\pdfoutput=1
\documentclass[10pt,conference,letterpaper]{IEEEtran}

\usepackage[protrusion=true,expansion=true]{microtype}
\usepackage{alltt}
\usepackage{amsmath}
\usepackage{array}
\usepackage{bussproofs}
\usepackage{comment}
\usepackage{floatrow}
\usepackage{graphicx}
\usepackage{subcaption}
\usepackage{tabularx}
\usepackage{times}
\usepackage{xspace}
\usepackage{flushend}
\def\lang{Bloom\xspace}
\def\blazes{\textsc{Blazes}\xspace}

\title{Blazes: Coordination Analysis for Distributed Programs}
\author{%
{Peter Alvaro{\small $~^{\#1}$}, Neil Conway{\small $~^{\#2}$}, Joseph M. Hellerstein{\small $~^{\#3}$},
David Maier{\small $~^{*4}$}
}%
\vspace{1.6mm}\\
\fontsize{10}{10}\selectfont\itshape
$^{\#}$\,UC Berkeley\\
\fontsize{9}{9}\selectfont\ttfamily\upshape
$^{1}$\,palvaro@cs.berkeley.edu\\
$^{2}$\,nrc@cs.berkeley.edu\\
$^{3}$\,hellerstein@cs.berkeley.edu%
\vspace{1.2mm}\\
\fontsize{10}{10}\selectfont\rmfamily\itshape
$^{*}$\,Portland State University\\
\fontsize{9}{9}\selectfont\ttfamily\upshape
$^{4}$\,maier@cs.pdx.edu
}
\begin{document}
\maketitle
\begin{abstract} 
  Distributed consistency is perhaps the most discussed topic
  in distributed systems today.
  Coordination protocols can ensure consistency, but
  in practice they cause undesirable performance unless used judiciously.
  Scalable distributed architectures avoid coordination
  whenever possible, but under-coordinated systems can exhibit
  behavioral anomalies under fault, which are often extremely difficult to 
  debug.
  This raises significant challenges for distributed system architects and 
  developers.
  % This presents significant challenges to developers: too much consistency
  % negatively impacts performance and manageability; too little results in
  % behavioral anomalies under fault, which are often extremely difficult to debug.

  In this paper we present \blazes, a cross-platform program analysis framework
  that
  % Coordination protocols are widely used to ensure the consistency of
  % distributed state.  However, these protocols have significant costs in both
  % increased latency and reduced availability.  Hence, enabling programmers to
  % reason about the coordination requirements of their programs---and how to
  % achieve sufficient coordination at minimal cost---is an important practical
  % problem. In this paper, we present Blazes, a framework that
  (a) identifies
  program locations that require coordination to ensure consistent executions,
  and (b) automatically synthesizes application-specific coordination code
  % By
  % leveraging program semantics, developer annotations, and network topology,
  % \blazes can synthesize application-specific coordination schemes
  that can significantly outperform general-purpose techniques. We present two case
  studies, one using annotated programs in the Twitter Storm system, and
  another using the Bloom declarative language.

  % We present a case study
  % showing how \blazes successfully identifies coordination requirements and can
  % synthesize the necessary coordination, achieving significant performance
  % improvements over a state-of-the-art general-purpose coordination protocol.

\end{abstract}

\section{Introduction}

\begin{quote}
\emph{The first principle of successful scalability is to batter the consistency mechanisms down to a minimum.} \\-- James Hamilton, as transcribed in~\cite{ladis-report}.
\end{quote}
\begin{quote}
\emph{When your map or guidebook indicates one route, and the blazes show another, follow the blazes.} \\-- Appalachian trail conservancy~\cite{a-t}.

\end{quote}
\begin{comment}
\begin{quote}
    \emph{Too slow for Boylan, blazes Boylan, impatience Boylan, joggled the mare.} -- James Joyce.
\end{quote}
\end{comment}

% Remember the community outreach gambit?  
Over the past decade, the database research community has deliberately 
widened its focus to ``maximize impact \ldots{} across computing'' by 
exploring a variety of computer science challenges that seem well-suited 
to data management technologies~\cite{claremont}.  One promising dimension 
of this expanded agenda is the exploration of declarative languages for 
new platforms and application domains, including (among others) network 
protocols, machine learning, and cloud computing. 

Distributed systems and cloud computing offer particularly attractive 
opportunities for declarative languages, given their focus on data-centric 
applications.  Initial work in this domain largely focused on benefits of 
code simplicity, showing that declarative languages are a natural fit for 
specifying distributed systems internals and can yield code that is 
easier to maintain and extend~\cite{idodeclare,boomanalytics,cidr,Loo2006}.  While that 
work has the potential for long-term software engineering benefits, it 
provides little short-term incentive for distributed systems developers to 
switch from their familiar imperative languages and tools.
% , however poorly 
% suited they may be to the parallel, asynchronous, unpredictable realities 
% of cloud computing.

In this paper, we show how database technology can address a significantly more
urgent issue for distributed systems developers: the correctness and efficiency
of distributed consistency mechanisms for fault-tolerant services.  The need for
consistency, and methods for achieving it, have been the subject of extensive
debate in the practitioner and research
community~\cite{ladis-report,Helland2007,Brewer2012}. Coordination services for
distributed consistency, such as Chubby~\cite{chubby} and
Zookeeper~\cite{zookeeper}, are in wide use. At the same time, there have been
various efforts to address consistency in system-specific ways, including NoSQL
systems~\cite{nosql-survey}, Internet services
infrastructure~\cite{chubby,dynamo,spanner,megastore} and even large-scale
machine learning systems~\cite{GraphLab,Hogwild}. The reason for the interest is
clear: for many practitioners distributed consistency is the most critical issue
for system performance and manageability at scale~\cite{ladis-report}.

\subsection{Blazes}
Recent work has highlighted promising connections between distributed 
consistency and database theory surrounding 
monotonicity~\cite{cidr,imperative,relational-transducers,winmove}.  In this 
paper we move beyond theory and language design to develop practical 
techniques that have direct utility for popular distributed programming 
platforms like Twitter Storm~\cite{stormbook}, while providing even more 
power for the declarative languages like Bloom~\cite{bloom} that are being 
designed for the future.  
%Specifically, we introduce a cross-language framework called \blazes that builds on prior results to (a) \emph{analyze} application software for distributed consistency problems, and (b) automatically \emph{generate} application-aware coordination logic that prevents such anomalies---in many cases without resorting to distributed coordination services like Chubby or Zookeeper, or global protocols like Multipaxos~\cite{paxos}.

Specifically, we present \blazes, a cross-language analysis framework that
provides developers of distributed programs with judiciously chosen,
application-specific coordination code.  First, \blazes \emph{analyzes}
applications to identify code that may cause consistency anomalies.  \blazes'
analysis is based on a pattern of properties and composition: it begins with key
properties of individual software components, including order-sensitivity,
statefulness, and replication; it then reasons transitively about compositions
of these properties across dataflows that span components.  Second, \blazes
automatically \emph{generates} application-aware coordination code to prevent
consistency anomalies with a minimum of coordination.  The key intuition
exploited by \blazes is that even when components are order-sensitive, it is
often possible to avoid the cost of global ordering
without sacrificing consistency. In many cases, \blazes can ensure consistent
outcomes via a more efficient and manageable protocol of asynchronous
point-to-point communication between producers and consumers---called
\emph{sealing}---that indicates when partitions of a stream have stopped
changing. These partitions are identified and ``chased'' through a dataflow via
techniques from functional dependency analysis, another surprising application
of database theory to distributed consistency.

\begin{figure}
\includegraphics[width=1.0\linewidth]{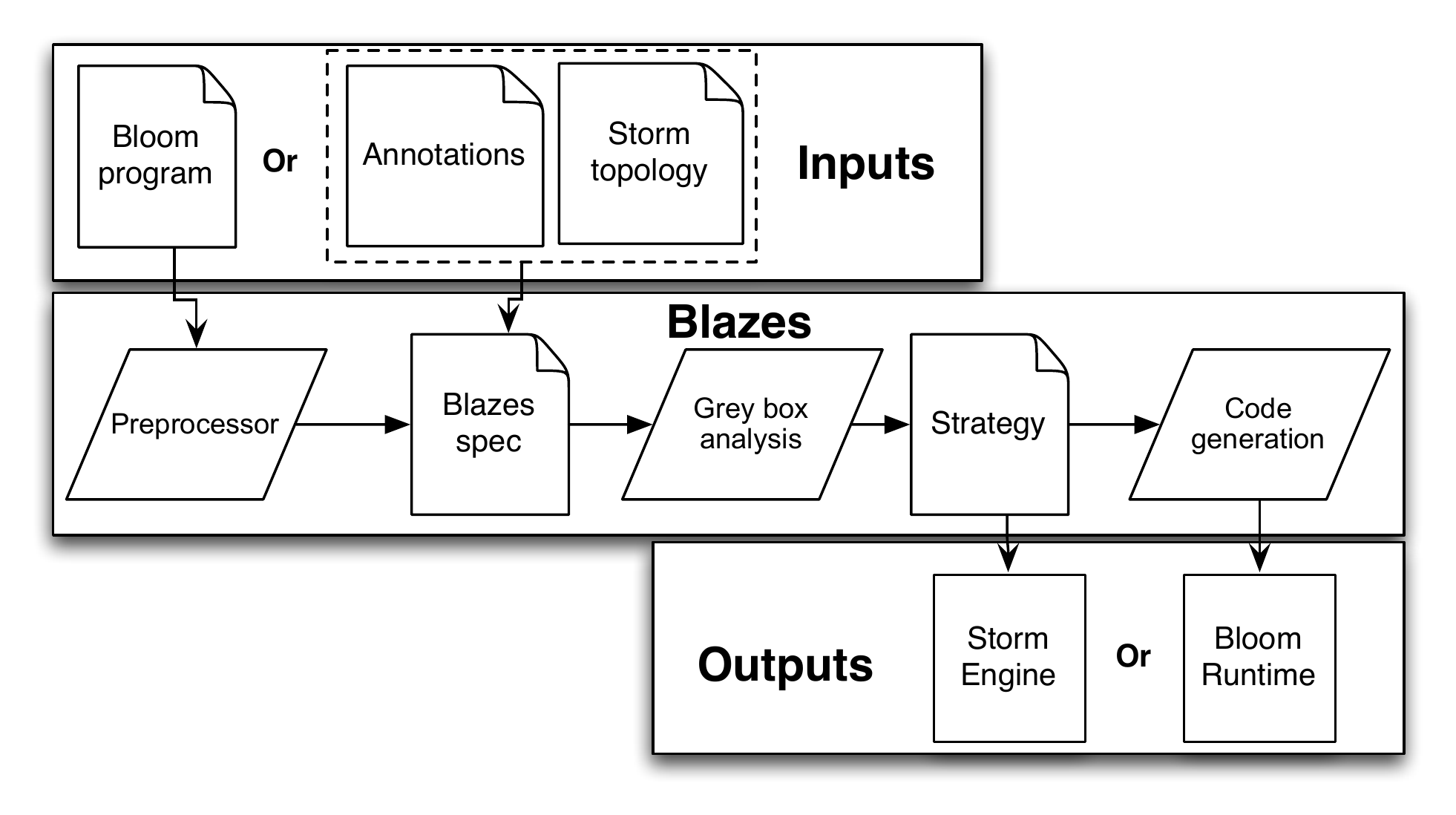}
\caption{\small The \blazes framework.  In the ``grey box'' system, programmers supply
  a configuration file representing an annotated dataflow.  In the ``white box''
  system, this file is automatically generated via static analysis.}
\label{fig:block}
\end{figure}

 The \blazes architecture is depicted in
  Figure~\ref{fig:block}.
 \blazes can be directly applied to existing programming platforms based on distributed stream or dataflow processing,
 including Twitter Storm~\cite{stormbook}, Apache S4~\cite{s4}, and Spark Streaming~\cite{spark-streaming}.
 Programmers of stream processing engines interact with \blazes in a ``grey box''
 manner: they provide simple semantic \emph{annotations} to the black-box
 components in their dataflows, and \blazes performs the analysis of all dataflow
 paths through the program.  
 \blazes can also
  take advantage of the richer analyzability of declarative languages like Bloom.
 Bloom programmers are freed from the need to supply
 annotations, since Bloom's language semantics allow \blazes to infer component
 properties automatically. 

We make the following contributions in this paper:
 \begin{itemize}
 \item \textbf{Consistency Anomalies and Properties.}  We present a spectrum of consistency anomalies that arise in distributed dataflows.  We identify key properties of both streams and components that affect consistency.  
 \item \textbf{Composition of Properties.} We show how to analyze the composition of consistency properties in complex programs via a term-rewriting technique over dataflow paths, which translates local component properties into end-to-end stream properties.
 \item \textbf{Custom Coordination Code.} We distinguish two alternative coordination strategies,
   \emph{ordering} and \emph{sealing}, and show how we can automatically generate application-aware coordination code that uses the cheaper sealing technique in many cases.
 \end{itemize}

 We conclude by evaluating the performance benefits offered by using
 \blazes as an alternative to generic, order-based coordination mechanisms available
 in both Storm and Bloom.%   Our experiments also suggest some important directions for future work that we discuss,
 % based on the interaction between data placement
 % on application-specific coordination strategies.

\subsection{Running Examples}
\label{sec:example}

\begin{figure}[t]
\centering
\includegraphics[width=0.75\linewidth]{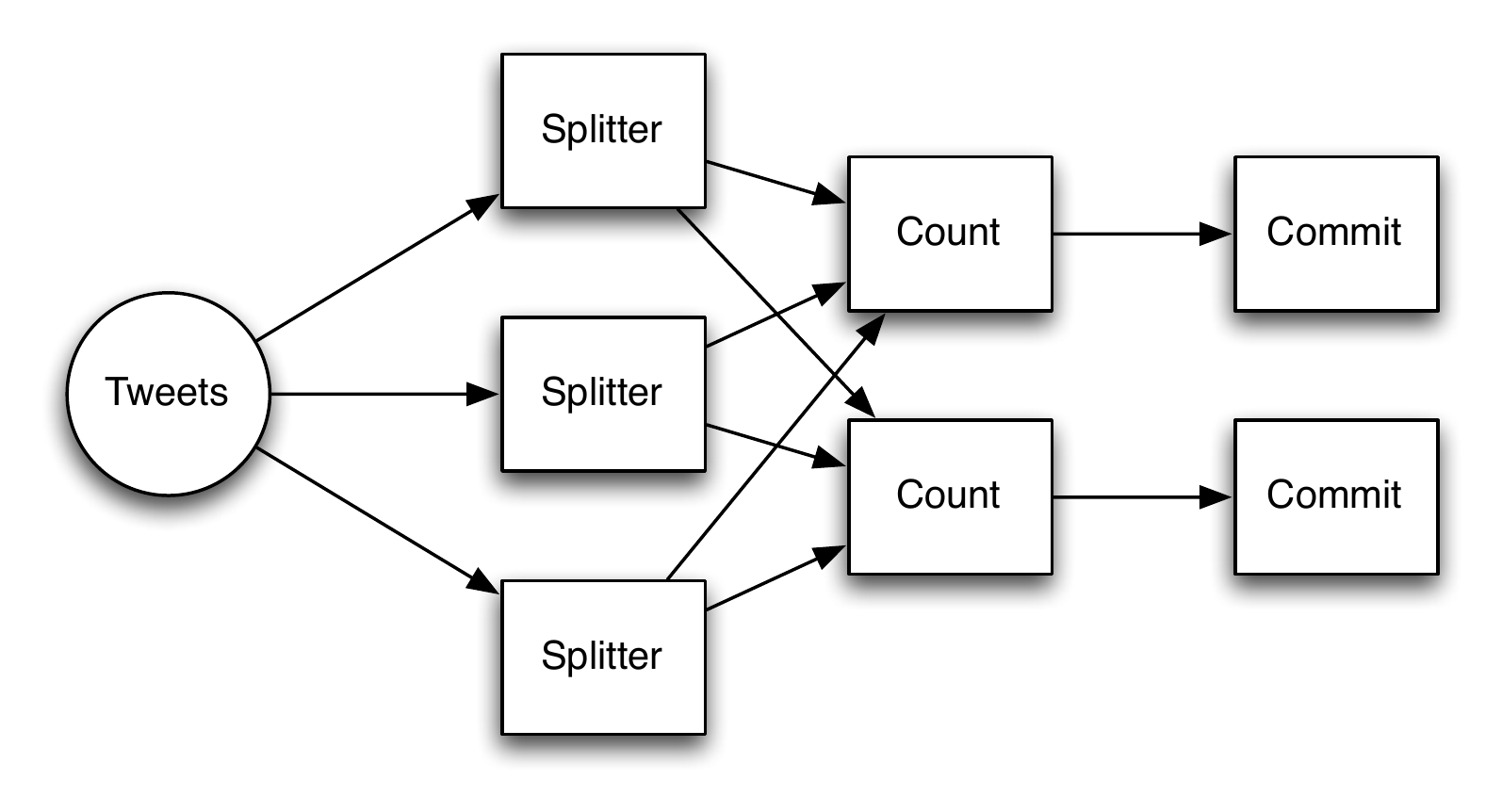}
\caption{\small Physical architecture of a Storm word count topology.}
\label{fig:storm_arch}
\end{figure}

\begin{figure}[t]
\centering
\includegraphics[width=0.8\linewidth]{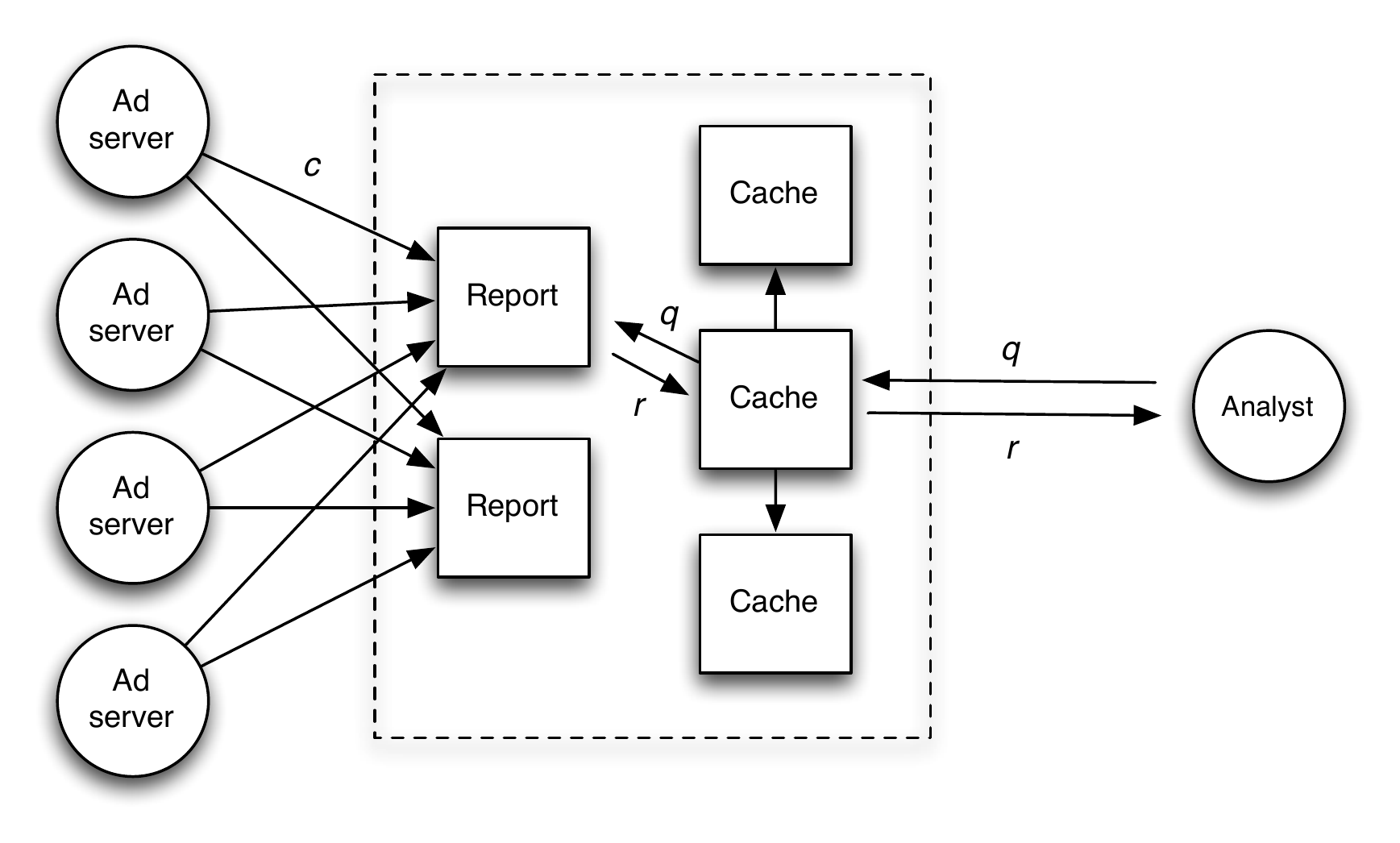}
\caption{\small Physical architecture of an ad-tracking network.}
\label{fig:ad_serving_arch}
\end{figure}

We consider two running examples: a streaming analytic query implemented using
the Storm stream processing system and a distributed ad-tracking network implemented
using the Bloom distributed programming language.

\vspace{0.5em}\noindent
\textbf{Streaming analytics with Storm:} Figure~\ref{fig:storm_arch} shows the
architecture of a Storm topology that computes a continuous word count over the
Twitter stream. 
Each ``tweet'' is associated with a numbered batch (the unit of replay) and is
sent to exactly one \texttt{Splitter} component---which divides tweets into their
constituent words---via random partitioning.
The words are hash partitioned to the 
\texttt{Count} component, which 
tallies the number of occurrences of each word in the
current batch. When a batch ends, the \texttt{Commit} component records the
batch number and frequency for each word in a backing store.

Storm ensures fault-tolerance via replay: if component instances fail or time out, 
stream sources redeliver their inputs.  
It is up to the programmer to ensure
that accurate counts are
committed to the store despite these at-least-once delivery semantics.  
One approach is to make the Storm topology
\emph{transactional}---i.e., one that processes tuples in atomic batches,
ensuring that certain components (called \emph{committers}) emit the batches in
a total order.  By recording the last successfully processed
batch identifier, a programmer may ensure at-most-once processing in the face of possible replay
by incurring the extra overhead of synchronizing the processing of batches.

Note that batches are independent in the word counting application; because the streaming query groups outputs
by batch id, there is no need to order batches with respect to each
other.
\blazes can aid a topology designer in avoiding unnecessarily conservative ordering
constraints, which (as we will see in Section~\ref{sec:eval}) results in up to a 
$3 \times$ improvement in throughput in our experiments.

\vspace{0.5em}\noindent
\textbf{Ad-tracking with Bloom:} Figure~\ref{fig:ad_serving_arch} depicts an
ad-tracking network, in which a collection of \emph{ad servers} deliver advertisements to users (not
shown) and send click logs (edges labeled ``\emph{c}'') to a set of
\emph{reporting server} replicas.  
% This architecture is based upon common practice that we observed at a large-scale internet
% search enterprise.
Reporting servers compute a continuous query;
analysts make requests (``\emph{q}'') for subsets of the query answer (e.g., by
visiting a ``dashboard'') and receive results via the stream labeled
``\emph{r}''.  To improve response times for common queries, 
a caching tier is interposed
between analysts and reporting servers. An analyst poses a request about a
particular ad to a cache server. If the cache contains an answer for the query,
it returns the answer directly. Otherwise, it forwards the request to a
reporting server; when a response is received, the cache updates its local state
and returns a response to the analyst. Asynchronously, it also sends the
response to the other caches.  The clickstream \emph{c}---each partition of which is
generated by a single ad
server---is sent to all reporting
servers; this improves fault tolerance and reduces query latency, because caches
can contact any reporting server.  Due to failure, retry and the interleaving of
messages from multiple ad servers, network delivery order is nondeterministic.  As
we shall see, different continuous queries have different sensitivities to
network nondeterminism.
\blazes can help determine how much coordination is required to
ensure that network behavior does not cause inconsistent results.

\section{System Model}
\label{sec:system-model}

The \blazes API is based on a simple ``black box'' model of component-based
distributed services.  
We use dataflow graphs~\cite{kpn} to represent
distributed services: nodes in the graph correspond to service components, which
expose input and output \emph{interfaces} that correspond to service calls or
other message events.  
%We can represent both the data- and control-flow of the
%ad server network using the dataflow diagram in Figure~\ref{fig:dataflow}.
While we focus our discussion on streaming analytics systems,
we can represent both the data- and control-flow of arbitrary distributed systems
using this dataflow model.

\begin{figure}[t]
\centering
\includegraphics[width=0.8\linewidth]{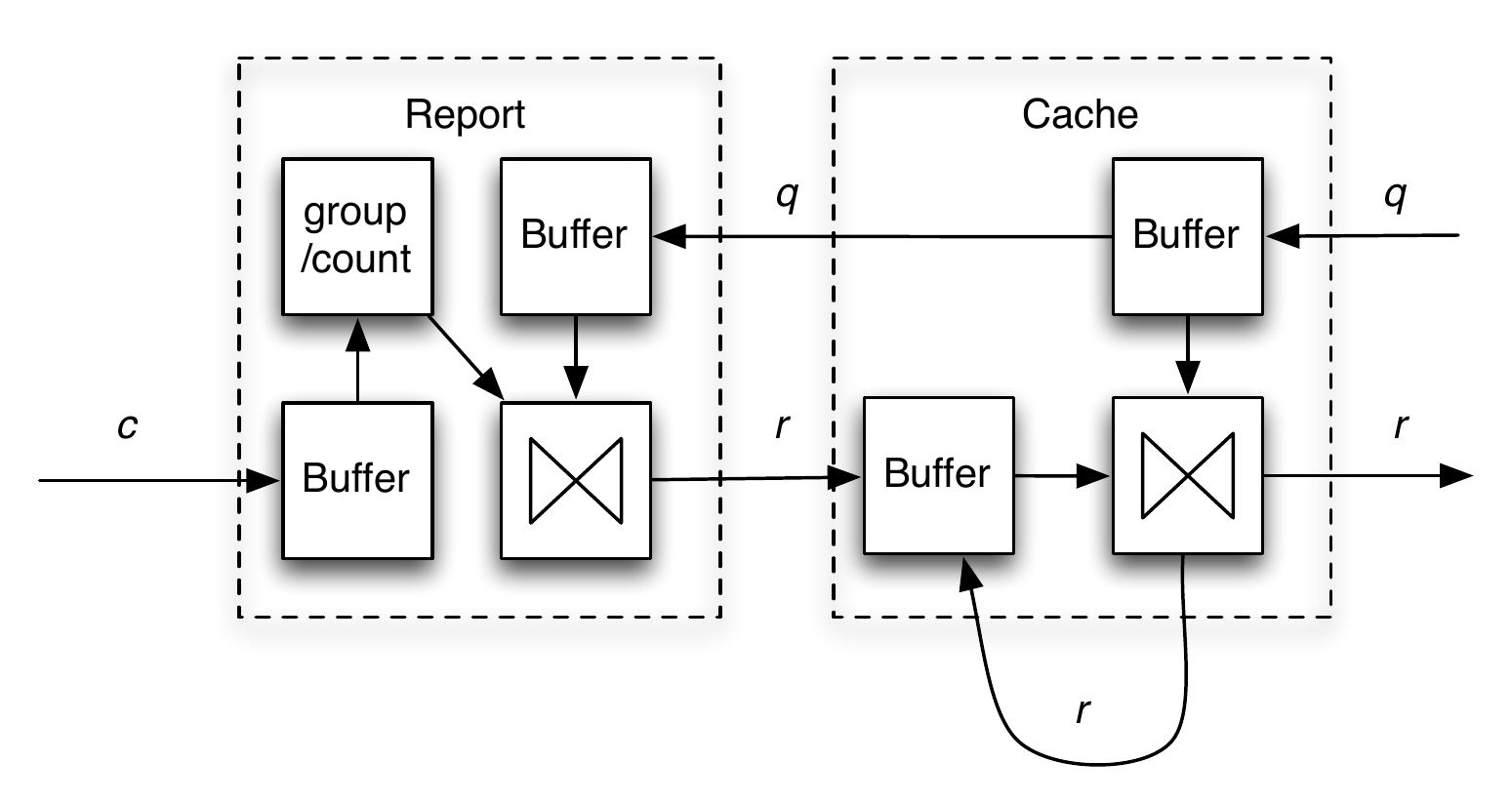}

\caption{\small Dataflow representations of the ad-tracking network's \texttt{Report} and \texttt{Cache} components.}
\label{fig:dataflow}
\end{figure}

\begin{comment}
The \emph{logical dataflow} in Figure~\ref{fig:dataflow} captures the \emph{software architecture}
of the ad tracking network, describing how components interact via API calls.
By contrast, the \emph{physical dataflow} shown in Figure~\ref{fig:ad_serving_arch} captures the
\emph{system architecture}, mapping software components to the
physical resources on which they will execute.
\end{comment}
A \emph{logical dataflow} (e.g., the representation of the ad tracking network depicted in Figure~\ref{fig:dataflow})
captures a \emph{software architecture}, describing how components interact via API calls.
By contrast, a \emph{physical dataflow} (like the one shown in Figure~\ref{fig:ad_serving_arch}) extends a software
architecture into a \emph{system architecture}, mapping software components to the
physical resources on which they will execute.
We choose to focus our analysis on logical dataflows,
which abstract away details like the multiplicity of physical resources but are
sufficient---when properly annotated---to characterize the consistency semantics of distributed services.

\subsection{Components and Streams}
A \emph{component} is a logical unit of computation and storage, processing
streams of inputs and producing streams of outputs over time. Components are
connected by \emph{streams}, which are unbounded, unordered~\cite{oop}
collections of messages.  A stream associates an output interface of one
component with an input interface of another. To reason about the behavior of a
component, we consider all the \emph{paths} that connect its inputs and
outputs. For example, the reporting server (\texttt{Report} in
Figure~\ref{fig:dataflow}) has two input streams, and hence defines two possible
dataflow paths (from \emph{c} to \emph{r} and from \emph{q} to \emph{r}).  We
assume that components are deterministic: two instances of a given component that receive the same
inputs in the same order produce the same outputs and reach the same state.

A \emph{component instance} is a binding between a component and a physical
resource---with its own clock and (potentially mutable) state---on which the
component executes.  In the ad system, the reporting server is a single logical
component in Figure~\ref{fig:dataflow}, but corresponds to two distinct
(replicated) component instances in Figure~\ref{fig:ad_serving_arch}.
Similarly, we differentiate between logical streams (which characterize the
types of the messages that flow between components) and \emph{stream instances},
which correspond to physical channels between component instances.  Individual
components may execute on different machines as separate component instances,
consuming stream instances with potentially different contents and orderings.

While streams are unbounded, in practice they are often divided into
batches~\cite{stormbook,spark-streaming,magda-survey} to enable replay-based
fault-tolerance.  \emph{Runs} are (possibly repeated) executions over finite
stream batches.

A stream producer can optionally embed \emph{punctuations}~\cite{punctuations}
into the stream.  A punctuation guarantees that the producer will generate no
more messages within a particular logical partition of the stream.  For example,
in Figure~\ref{fig:ad_serving_arch}, an ad server might indicate that it will
henceforth produce no new records for a particular time window or advertising
campaign via the \emph{c} stream.  In Section~\ref{sec:coordination}, we show
how punctuations can enable efficient, localized coordination strategies based
on \emph{sealing}.
% Punctuations must contain metadata describing the contents
% of the partition that they seal, because (given our conservative assumptions
% regarding stream order) a punctuation for a partition may arrive before some of
% the contents of that partition.

\begin{comment}
\subsection{Dataflows}

\paa{this section adds little and infuriates neil.  it is hereby marked for GC}

Connecting a collection of components and streams yields a logical dataflow
graph.  Because components are defined recursively, we can treat any such
graph as a component; its input (respectively, output) interfaces correspond
to the unconnected input (output) streams.  For example, in Figure~\ref{fig:dataflow},
the diagram on the left abstracts the logic
of the reporting server into a single functional unit, while the latter ``unpacks'' the reporting logic (\texttt{Report}) into a
lower-level operator
dataflow reminiscent of a database query plan.  

In a particular system deployment, a logical dataflow of components and streams corresponds to an unbounded number of \emph{physical dataflows},
representing different one-to-many mappings of components (streams) to component (stream) instances.  
\end{comment}

\section{Dataflow Consistency}
\label{sec:properties}

In this section, we develop consistency criteria and mechanisms appropriate to distributed, fault-tolerant
dataflows.  We begin by describing undesirable behaviors that can arise 
due to the interaction between nondeterministic message orders and fault-tolerance mechanisms.
We review common strategies for preventing such anomalies by exploiting semantic
properties of components (Section~\ref{sec:calm}) or by enforcing constraints on
message delivery (Section~\ref{sec:mechanism}).
We then 
generalize delivery mechanisms
into two classes: message \emph{ordering} and
partition \emph{sealing}.
Finally, we consider a collection of queries that we could install at the reporting server in the 
ad tracking example presented in Section~\ref{sec:example}.
We show how slight differences in the queries can lead to different distributed anomalies,
and how practical variants of the ordering and sealing strategies can be used to prevent these anomalies.

\subsection{Anomalies}
\label{sec:anomalies}

\begin{figure}
\includegraphics[width=1.0\linewidth]{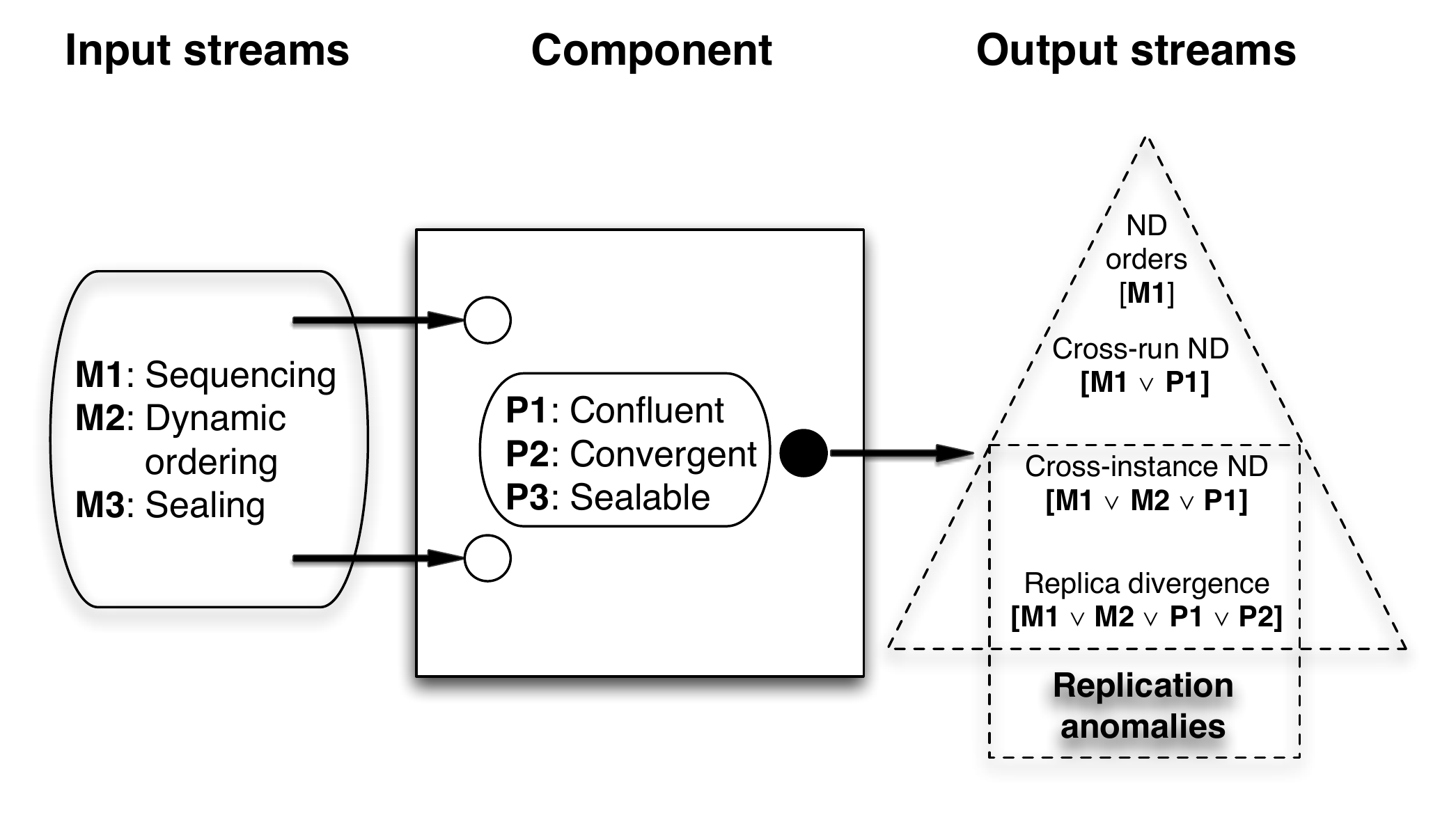}
\caption{\small The relationship between potential stream anomalies (right) and remediation strategies based on
component properties (center) and delivery mechanisms (left).  
For each anomaly, we list the properties and mechanisms that \emph{prevent} it.
For example, convergent components (\textbf{P2}) prevent only replica divergence
 while a dynamic message ordering mechanism (\textbf{M2}) prevents all replication anomalies.}
\label{fig:anomalies}
\end{figure}

Nondeterministic messaging interacts with fault-tolerance mechanisms in subtle
ways.  Two standard schemes exist for fault-tolerant dataflows:
\emph{replication} (used in the ad reporting system described in
Section~\ref{sec:example}) and \emph{replay} (employed by Storm and
Spark)~\cite{magda-survey}.  
%Both mechanisms use \emph{redundant} state and computation to guard against failures,
%allowing systems to recover from the loss of components or messages.
Both mechanisms allow systems to recover from the failure of components or the loss of messages
via \emph{redundancy} of state and computation.
Unfortunately, redundancy brings with it a need to consider issues of \emph{consistency},
because nondeterministic message orders
can lead to disagreement regarding stream contents among replicas or across replays.
This disagreement
undermines the transparency that fault tolerance mechanisms
are meant to achieve, giving rise to anomalies that are difficult to debug.

Figure~\ref{fig:anomalies}---which depicts a component with two input streams and a single output stream---captures 
the relationship between delivery mechanisms, component properties
and output stream anomalies.
The spectrum of behaviors that can arise as a result of
nondeterministic message ordering are listed on the right side of the figure.
%A component output stream is anomaly-free if
%its contents and order are uniquely determined by its inputs.  
Because
it is difficult to control the order in which a component's inputs appear,
the first (and least severe) anomaly is nondeterministic orderings
of (otherwise deterministic) output contents (\textbf{Async}).  
In this paper, we focus on the three remaining classes of anomalies, all of which
have direct consequences on the fault-tolerance mechanism:

\begin{enumerate}
  \item \emph{Cross-run nondeterminism} (\textbf{Run}), in which a component
    produces different output stream \emph{contents} in different
    runs over the same inputs.
    %Systems that do not exhibit cross-run nondeterminism are \emph{replayable}, and support efficient replay-based fault-tolerance.
    Systems that exhibit cross-run nondeterminism do not support replay-based fault-tolerance.
    For obvious reasons, nondeterminism across runs makes such systems difficult to test and debug.
  \item \emph{Cross-instance nondeterminism} (\textbf{Inst}), in which replicated instances of the same components produce
    different output contents in the \emph{same} execution over the same inputs.
    Cross-instance nondeterminism can lead to inconsistencies across queries.
  \item \emph{Replica divergence} (\textbf{Diverge}),
    in which the state of multiple replicas becomes permanently inconsistent.
    Some services may tolerate transient disagreement between streams (e.g., for
    streams corresponding to the results of read-only queries), but permanent
    replica divergence is never desirable.

\end{enumerate}

% All anomalies are \emph{witnesses} of asynchronous execution; nondeterminism in message ordering has ``leaked''
% into program outputs.

\begin{figure*}[t]
\begin{center}
\begin{small}
\begin{tabular}{|l|l|}
\hline
\emph{Name} & \emph{Continuous Query}  \\ \hline
THRESH &  \texttt{select id from clicks group by id having count(*) > 1000} \\ \hline
POOR &  \texttt{select id from clicks group by id having count(*) < 100} \\ \hline
WINDOW &  \texttt{select window, id from clicks group by window, id having count(*) < 100} \\ \hline
CAMPAIGN &  \texttt{select campaign, id from clicks group by campaign, id having count(*) < 100} \\ \hline
\end{tabular}
\end{small}
\end{center}
\caption{\small Reporting server queries (shown in SQL syntax for familiarity).}
\label{fig:sql}
\end{figure*}

\subsection{Monotonicity, confluence and convergence}
\label{sec:calm}

%Fortunately, a class of programs can be proven immune to the consistency
%anomalies described above.  
Distributed programs that produce the same outcome
for all message delivery orders
exhibit none of the anomalies listed in Section~\ref{sec:anomalies},
regardless of the choice of fault-tolerance or delivery mechanisms.
In recent work, we proposed the \emph{CALM
  theorem}, which observes that 
%\emph{monotonic} programs produce consistent
%results regardless of nondeterminism in delivery
% orders~\cite{cidr,imperative,relational-transducers}. 
programs expressed in \emph{monotonic} logic produce deterministic results
despite nondeterminism in delivery orders~\cite{cidr,imperative,relational-transducers}.
Intuitively, monotonic
programs compute a continually growing result, never retracting an earlier
output given new inputs.  Hence replicas running monotonic code always
eventually agree, and replaying monotonic code produces the same result in every
run.

We call a dataflow component \emph{confluent} if it produces the same \emph{set} of outputs
for all \emph{orderings} of its inputs.  At any time, the output of a confluent 
component (and any redundant copies of that component) 
is a subset of the unique, ``final'' output.  
Confluent components never exhibit
any of the three dataflow anomalies listed above.  Confluence is a property of the behavior of
components---monotonicity (a property of program logic) is a sufficient condition
for confluence.

%Distributed systems commonly adopt a storage-centric notion of consistency.
%Replicated storage is \emph{eventually consistent} or \emph{convergent} if, when
%all messages have been delivered, all replicas agree on the set of stored
%values~\cite{vogels-ec}.  Convergent components never exhibit replica divergence.
Confluence is similar to the notion of replica \emph{convergence} common in distributed systems.
A system is convergent or ``eventually consistent'' if, when all messages have been delivered,
all replicas agree on the set of stored values~\cite{vogels-ec}.  Convergent components never exhibit
replica divergence.  Convergence is a local guarantee
about component \emph{state}; by contrast, confluence provides guarantees about component \emph{outputs},
which (because they become the inputs to downstream components) compose into global guarantees about dataflows.

Confluence implies convergence but the converse does not hold. Convergent
replicated components are guaranteed to eventually reach the same state, but
this final state may not be uniquely determined by component inputs.  As
Figure~\ref{fig:anomalies} indicates, convergent components allow cross-instance
nondeterminism, which can occur when reading ``snapshots'' of the convergent
state while it is still changing.  Consider what happens when the read-only
outputs of a convergent component (e.g., GETs posed to a key/value store) flow
into a replicated stateful component (e.g., a replicated cache).  If the caches
record different stream contents, the result is replica divergence.

%Storage-centric consistency criteria focus on the properties of data at rest;
%reasoning about the overall properties
%of a composed dataflow requires following the data as it moves.

\subsection{Coordination Strategies}
\label{sec:mechanism}

Confluent components produce deterministic outputs and convergent replicated
state.  How can we achieve these properties for components that are not
confluent?  We assume that components are deterministic, so we can prevent
inconsistent outputs within or across program runs simply by removing the
nondeterminism from component input orderings.  Two extreme approaches include
(a) establishing a single total order in which all instances of a given
component receive messages (a \emph{sequencing} strategy) and (b) disallowing
components from producing outputs until all of their inputs have arrived (a
\emph{sealing} strategy).  The former---which enforces a total order of
inputs---resembles state machine replication from the distributed systems
literature~\cite{statemachine}, a technique for implementing consistent
replicated services.  The latter---which instead controls the order of
evaluation at a coarse grain---resembles stratified evaluation of logic
programs~\cite{ullmanbook} in the database literature, as well as barrier synchronization
mechanisms used in systems like MapReduce~\cite{mapreduce}.

Both strategies lead to ``eventually consistent'' program outcomes---if we wait
long enough, we get a unique output for a given input.  Unfortunately, neither
leads directly to a practical coordination implementation.  We cannot in general
preordain a total order over all messages to be respected in all executions.
Nor can we wait for streams to stop producing inputs, as streams are unbounded.

Fortunately, both coordination strategies have a dynamic variant that allows
systems to make incremental progress over time.  To prevent
replica divergence, it is sufficient to use a dynamic ordering service
(e.g., Paxos) that decides a global order of messages \emph{within a particular
run}.  As Figure~\ref{fig:anomalies} shows, a nondeterministic choice of message 
ordering can prevent cross-instance
nondeterminism but not cross-run nondeterminism since the choice is
dependent on arrival orders at the coordination service.  Similarly, strategies
based on sealing inputs can be applied to infinite streams as long as the
streams can be partitioned into finite
partitions that exhibit temporal locality, like windows with ``slack''~\cite{aurora}.
Sealing strategies---applicable when input stream partitioning is \emph{compatible} with
component semantics---can rule out all nondeterminism 
anomalies.\footnote{That is, \textbf{M3} and \textbf{P3} together are semantically equivalent to \textbf{P1} in Figure~\ref{fig:anomalies}.  
The notion of compatibility is defined in Section~\ref{sec:analysis}.}
Note that sealing is significantly less constrained than
ordering: it enforces an output barrier per partition, but allows asynchrony
both in the arrival of a batch's inputs and in interleaving across batches.

\subsection{Example Queries}
\label{sec:queries}

The ad reporting system presented in Section~\ref{sec:example} involves a collection of 
components interacting in a dataflow graph.
In this section, we focus on the \texttt{Report} component, which accumulates 
click logs and continually evaluates a standing query against them.
Figure~\ref{fig:sql} presents a variety of simple queries that we might install at the reporting server; perhaps surprisingly,
these queries have substantially different coordination requirements if we demand that they return deterministic answers.
%In Section~\ref{sec:case}, we will use \blazes to determine these requirements
%by treating each query as a separate instance of the \texttt{Report}
%component.
%While the complete ad reporting network we evaluate in
%Section~\ref{sec:eval} is implemented in the Bloom language, here we 
%consider the reporting queries in SQL syntax for familiarity.

We consider first a threshold query
\emph{THRESH}, which computes the unique identifiers of any ads that have at
least 1000 impressions.  
%Although the click messages may arrive in different
%orders at different replicas or in different executions, \emph{THRESH} returns
%results only when a count of messages exceeds a threshold.  
\emph{THRESH} is confluent: we expect it to produce a deterministic result set without need for coordination, 
since the value of the count monotonically increases in a manner insensitive to 
message arrival order~\cite{lattice}.

By contrast, consider a ``poor performers'' query: \emph{POOR}
returns the IDs of ads
that have fewer than one hundred clicks (this might be used to recommend such
ads for removal from subsequent campaigns).
\emph{POOR} is nonmonotonic: as more clicks are observed, the set of poorly
performing ads might shrink---and because it ranges over the entire
clickstream, we would have to wait until there were no more log messages to
ensure a unique query answer.  Allowing \emph{POOR} to emit results ``early'' based on a
nondeterministic event, like a timer or request arrival, is potentially
dangerous; multiple reporting server replicas
could report different answers in the same execution.
To avoid such anomalies, replicas could remain in sync by coordinating
to enforce a global message delivery order.
Unfortunately,  
this approach incurs significant latency and availability costs.

In practice, streaming query systems often address the
problem of blocking operators via \emph{windowing}, which constrains blocking queries to operate over bounded inputs~\cite{aurora,stream,telegraphcq}.
%We can exploit the semantics of streaming windows to guarantee deterministic query results.  
If the poor performers threshold test is
\emph{scoped} to apply only to individual windows (e.g., by including the window
name in the grouping clause), then ensuring deterministic
results is simply a matter of blocking until there are no more log messages \emph{for that window}.
Query \emph{WINDOW} returns, for each one
hour window, those advertisement identifiers that
have fewer than 100 clicks within that window.  

%The windowing strategy---a special case of \emph{sealing}---ensures deterministic results by delaying the nonmonotonic query from
%processing a logical partition of the input stream until its contents are completely determined.
%This sealing technique may also be applied to partitions that are not explicitly temporal.
The windowing strategy is a special case of the more general technique of \emph{sealing}, which 
may also be applied to partitions that are not explicitly temporal.
For example, it is common practice to associate a collection of ads with a ``campaign,'' or a grouping of advertisements
with a similar theme.  Campaigns may have different lengths, and may
overlap or contain other campaigns.
Nevertheless, given a punctuation indicating the termination of a campaign, the nonmonotonic query \emph{CAMPAIGN}
can produce deterministic outputs.

\section{Annotated Dataflow Graphs}
\label{sec:grey}

So far, we have focused on the consistency anomalies that can affect individual
``black box'' components. In this section, we extend our discussion 
by presenting a \emph{grey box} model in which programmers provide
simple annotations about the semantic properties of components. 
In Section~\ref{sec:analysis}, we show
how \blazes can use these annotations to automatically derive the consistency
properties of entire dataflow graphs.

\subsection{Annotations and Labels}

\begin{figure}[t]
\begin{center}
\begin{tabular}{|l|l|l|l|}
\hline
\emph{Severity} & \emph{Label} & \emph{Confluent} & \emph{Stateless} \\ \hline
1 & $CR$ & \textbf{X} & \textbf{X} \\ \hline
2 & $CW$ & \textbf{X} & \\ \hline
3 & $OR_{gate}$ & & \textbf{X} \\ \hline
4 & $OW_{gate}$ & &\\ \hline
\end{tabular} 
\end{center}
\caption{\small The \textbf{C.O.W.R.} component annotations.  A component path is
either \textbf{C}onfluent or \textbf{O}rder-sensitive, and either changes
component state (a \textbf{W}rite path) or does not (a \textbf{R}ead-only path).
Component paths with higher \emph{severity} annotations can produce more stream
anomalies.
}
\label{fig:component-annotations}
\end{figure}

\begin{figure}[t]
\begin{small}
\begin{tabular}{|l|l|l|l|l|l|}
\hline
\emph{S} & \emph{Label} & \emph{ND} & \emph{ND} & \emph{Transient} & \emph{Persistent} \\ 
 & & \emph{order} & \emph{contents} & \emph{replica} & \emph{replica} \\ 
 & &  &  & \emph{divergence} & \emph{divergence} \\ \hline

0 & \textbf{NDRead$_{gate}$} & \textbf{X} & \textbf{X} & & \\ \hline
0 & \textbf{Taint} & \textbf{X} & \textbf{X} & & \\ \hline
\hline
1 & \textbf{Seal$_{key}$} & \textbf{X} & & & \\ \hline
2 & \textbf{Async} & \textbf{X} & & & \\ \hline
3 & \textbf{Run} & \textbf{X} & \textbf{X} & & \\ \hline
4 & \textbf{Inst} & \textbf{X} & \textbf{X} & \textbf{X} & \\ \hline
5 & \textbf{Diverge} & \textbf{X} & \textbf{X} & \textbf{X} & \textbf{X} \\ \hline
\end{tabular}
\caption{Stream labels, ranked by severity (S).  \textbf{NDRead$_{gate}$} and \textbf{Taint} are internal labels, used by the analysis system but never output. \textbf{Run, Inst} and \textbf{Diverge} correspond to the stream anomalies enumerated in 
Section~\ref{sec:anomalies}: cross-run nondeterminism, cross-instance nondeterminism and replica divergence, respectively.}
\label{fig:stream-labels}

\end{small}
\end{figure}

In this section, we describe a language of \emph{annotations} and \emph{labels}
that enriches the ``black box'' model (Section~\ref{sec:system-model}) with
additional semantic information. Programmers supply annotations about paths
through components and about input streams; using this information, \blazes
derives labels for each component's output streams.

\subsubsection{Component Annotations}
\label{sec:comp-annotations}

\blazes provides a small, intuitive set of annotations that capture
component properties relevant to
stream consistency.  A review of the implementation or analysis of a component's
input/output behavior should be sufficient to choose an appropriate 
annotation.
Figure~\ref{fig:component-annotations} lists the component annotations supported
by \blazes. Each annotation applies to a path from an input interface to an
output interface; if a component has multiple input or output interfaces, each
path can have a different annotation.

The $CR$ annotation indicates that a path through a component is confluent and
stateless; that is, it produces deterministic output regardless of its input
order, and its inputs do not modify the component's state. $CW$ denotes a path
that is confluent and stateful.

The annotations $OR_{gate}$ and $OW_{gate}$ denote non-confluent paths that are stateless or
stateful, respectively. The $gate$ subscript is a set of attribute names that
indicates the partitions of the input streams over which the non-confluent component
operates. This annotation allows \blazes to determine whether an input stream
containing end-of-partition punctuations can produce deterministic executions
without using global coordination. Supplying ${gate}$ is optional; if the programmer
does not know 
%of any punctuation information, 
the partitions over which the component path operates,
the annotations $OR_*$ and $OW_*$
indicate that each record belongs to a different partition.

Consider a reporting server component implementing the query \emph{WINDOW}.  
%It
%consumes a stream of log entries and a stream of queries via its two input
%interfaces.  
When it receives a request referencing a particular advertisement
and window, it returns a response if the advertisement has fewer than 1000
clicks \emph{within that window}.  An appropriate label for the
 path from request inputs
to outputs as $OR_{id, window}$---a stateless order-sensitive path operating over
partitions with composite key \emph{id,window}.  Requests do not affect the
internal state of the component, but they do return potentially nondeterministic
results that depend on the outcomes of races between queries and click records.
Note however that if we
were to delay the results of queries until we were certain that there would be
no new records for a particular advertisement \emph{or} a particular
window,\footnote{This rules out races by ensuring 
(without enforcing an ordering on message delivery)
that the query comes
\emph{after} all relevant click records.} the output would be deterministic.
Hence \emph{WINDOW} is compatible with click streams partitioned  
(and emitting appropriate punctuations) on \emph{id} or \emph{window}.

\subsubsection{Stream Annotations}
Programmers can also supply optional annotations to describe the semantics of
streams. The \textbf{Seal$_{key}$} annotation means that the stream is
\emph{punctuated} on the subset $key$ of the stream's attributes---that is,
the stream contains punctuations on $key$, and there is at least one punctuation
corresponding to every stream record.
For example, a
stream representing messages between a client and server might have the label
\textbf{Seal$_{session}$}, to indicate that clients will send messages
indicating that sessions are complete.  To ensure progress, there must be
a punctuation for every session identifier that appears in any message.

Programmers can use the \textbf{Rep} annotation to indicate that a stream is
\emph{replicated}. A replicated stream connects a producer component instance (or instances)
to more than one consumer component instance, and produces the same contents for all stream instances
(unlike, for example, a partitioned stream).
The \textbf{Rep} annotation carries semantic information both about
expected execution \emph{topology} and \emph{programmer intent}, which \blazes
uses to determine when nondeterministic stream contents can lead to replica divergence.
\textbf{Rep} is an optional Boolean flag that may be combined
with other annotations and labels.

\subsubsection{Derived Stream Labels}
Given an annotated component with labeled input streams, \blazes can derive a
label for each of its output streams. Figure~\ref{fig:stream-labels} lists the
derived stream labels---each corresponds to a class of anomalies that may
occur 
%in the instances associated with streams having that label.  
in a given stream instance.
The label
\textbf{Async} corresponds to streams with deterministic contents
whose order may differ on different
executions or different stream instances.
\textbf{Async} is conservatively applied as the default
label; in general, we assume that communication between components is
asynchronous.

Streams labeled \textbf{Run} may exhibit cross-run nondeterminism, having
different contents in different runs.  Those labeled \textbf{Inst} may also
exhibit cross-instance nondeterminism on different replicas within a single run.  Finally, streams
labeled \textbf{Diverge} may exhibit persistent replica
divergence.

\section{Coordination Analysis and Synthesis}
\label{sec:analysis}

\blazes uses component and 
stream annotations to determine
if a given dataflow is guaranteed to produce deterministic outcomes; if it cannot
make this guarantee, it augments the program with coordination code.
In this section, we describe the program analysis and synthesis process.

\subsection{Analysis}

To derive labels for the output streams in a dataflow graph, \blazes starts by
enumerating all paths between pairs of sources and sinks.  To
rule out infinite paths, it reduces each cycle in the
graph to a single node with a collapsed label by selecting the label of highest
severity among the cycle members.  
\footnote{Note that in the ad-tracking network dataflow shown in Figure~\ref{fig:dataflow},
\texttt{Cache} participates in a cycle (the self-edge, corresponding to
communication with other cache instances), but \texttt{Cache} and
\texttt{Report} form no cycle, because \texttt{Cache} provides no path
from $r$ to $q$.}

For each component whose input streams are labeled (beginning with
the components with unconnected inputs), \blazes first performs an
\emph{inference} step, shown in Figure~\ref{fig:inference}, for every path
through the component.
When it has done so,
each of the output interfaces of the component is associated with a set of
derived stream labels (at least one for each distinct path from an input interface,
as well as the intermediate labels introduced by the \emph{inference}
rules).
\blazes then performs the second analysis step, the \emph{reconciliation} procedure
(described in Figure~\ref{fig:reconciliation}), which may add additional labels.
Finally, the labels for each output interface are merged into a single label.
This output stream becomes an
input stream of the next component in the dataflow, and so on
until all output streams are labeled.

\subsubsection{Transitivity of seals}

For non-confluent components with sealed input streams, the inference procedure must test whether the component preserves the independence of the sealed partitions---if it does, \blazes can ensure deterministic outcomes by delaying processing of partitions until when their complete contents are known.
For example, given the queries in Figure~\ref{fig:sql},
an input stream sealed on \emph{campaign} is only compatible with the query \emph{CAMPAIGN}---all
other queries combine the results from multiple campaigns into their answer, and may produce different
outputs given different message and punctuation orderings.

A stream sealed on key $key$ is compatible with a component with annotation $OR_{gate}$ or $OW_{gate}$ if
\emph{at least one} of the
attributes in $gate$ is \emph{injectively} determined by \emph{all} of the attributes in $key$.
For example, a company
name may functionally determine their stock symbol and the location of their
headquarters; when the company name Yahoo! is sealed their stock symbol
\texttt{YHOO} is implicitly sealed as well, but the city of Sunnyvale is not.
A trivial (and
ubiquitous) example of an injective function between input and output attributes
is the identity function,
which is applied
whenever we project an attribute without transformation.

We define the predicate $\textit{injectivefd}(A, B)$, which holds for attribute sets $A$
and $B$ if $A \mapsto B$ ($A$ functionally determines $B$) via some injective (distinctness-preserving)
function.
Such functions preserve the property of sealing: if we have seen all of the
$A$s, then we have also seen all the $f(A)$ for some injective $f$.

We may now define the predicate \emph{compatible}:
\begin{small}
\begin{align*}
\text{compatible(partition, seal)} \equiv \exists \text{ } attr \subseteq \text{partition } | \text{ injectivefd}(\text{seal}, attr)
\end{align*}
\end{small}

The compatible predicate will allow the \emph{inference} and \emph{reconciliation} procedures to test whether a sealed input stream matches the implicit
partitioning of a component path annotated $OW_{gate}$ or $OR_{gate}$.
In the remainder of this section we describe the \emph{inference} and \emph{reconciliation} procedures in detail.

\subsubsection{Inference}

At each reduction step, we apply the rules in Figure~\ref{fig:inference} to
derive additional intermediate stream labels for a component path.
An intermediate stream label may be any of the labels in Figure~\ref{fig:stream-labels}.

\begin{figure}
\begin{small}

\begin{prooftree}
\AxiomC{\{\textbf{Async, Run}\}}
\AxiomC{OR$_{gate}$}
\LeftLabel{(1)}
\BinaryInfC{\textbf{NDRead$_{gate}$}}
\end{prooftree}

\mbox{
\begin{subfigure}{0.5\linewidth}

\begin{prooftree}
\AxiomC{\{\textbf{Async, Run}\}}
\AxiomC{OW$_{gate}$}
\LeftLabel{(2)}
\BinaryInfC{\textbf{Taint}}
\end{prooftree}

\end{subfigure}
\quad

\begin{subfigure}{0.5\linewidth}
\begin{prooftree}
\AxiomC{\textbf{Inst}}
\AxiomC{CW, OW$_{gate}$}
\LeftLabel{(3)}
\BinaryInfC{\textbf{Taint}}
\end{prooftree}
\end{subfigure}
}

\begin{prooftree}
\AxiomC{\textbf{Seal$_{key}$}}
\AxiomC{OW$_{gate}$}
\AxiomC{$\lnot$ compatible($gate, key$)}
\LeftLabel{(4)}
\TrinaryInfC{\textbf{Taint}}
\end{prooftree}

\end{small}
\caption{Reduction rules for component paths.
Each rule takes an input stream label and a component annotation,
and produces a new (internal) stream label.
Rules may be read as implications: if the premises (expressions above the line)
hold, then the conclusion (below) should be added to the \texttt{Labels} list.
}
\label{fig:inference}
\end{figure}

Rules 1 and 2 of Figure~\ref{fig:inference} reflect the consequences of
providing nondeterministically ordered inputs to order-sensitive components.
\textbf{Taint} indicates that the internal state of the component may become
corrupted by unordered inputs.  \textbf{NDRead$_{gate}$} indicates that the
output stream may have transient nondeterministic contents.  Rule 3 captures the
interaction between cross-instance nondeterminism and replica divergence: transient
disagreement among replicated streams can lead to permanent replica divergence
if the streams modify component state downstream.
Rules 4 tests whether the predicate \emph{compatible} (defined in the previous
section) holds, in order to determine when sealed input streams are compatible
with stateful, non-confluent components.

When \emph{inference} completes, each output interface of the component is
associated with a list \texttt{Labels} of stream labels, containing all input stream
labels as well as any intermediate labels derived by inference rules.

\subsubsection{Reconciliation}

Given an output interface associated with a set of labels,
\blazes derives additional labels by using the \emph{reconciliation}
procedure shown in Figure~\ref{fig:reconciliation}.

If the \emph{inference} procedure has already determined that component state is tainted,
then the output stream may exhibit replica divergence (if the component
is replicated) and cross-run nondeterminism.
If $NDRead_{gate}$ (for some partition key $gate$) is among the stream labels,
the output interface may have nondeterministic contents given nondeterministic
input orders or interleavings with other component inputs, unless \emph{all} streams with which it can ``rendezvous''
are sealed on a compatible key.  If the component is replicated, nondeterministic outputs can lead
to cross-instance nondeterminism.

Once the internal labels have been dealt with, \blazes simply returns
the label in \texttt{Labels} of highest severity.

\subsubsection{Notation}

When describing trees of inferences, reconciliations and merges used to derive output stream labels,
we will use the following notation:

\begin{prooftree}
\AxiomC{SL$_1$}
\AxiomC{CA$_1$}
\LeftLabel{\small (R$_1)$}
\BinaryInfC{SL$_2$}
                              \AxiomC{SL$_3$}
                              \AxiomC{CA$_2$}
                              \LeftLabel{\small (R$_2)$}
                              \BinaryInfC{SL$_4$}
                                                            \AxiomC{[$\ldots]$}
        \LeftLabel{\small CN$_1$}
        \TrinaryInfC{SL$_5$}
\end{prooftree}

Here the \emph{SL} are stream labels, the \emph{CA} are component annotations,
\emph{R} is the inference rule applied, and \emph{CN} is the component name
whose outputs are combined by the merge procedure.\footnote{For ease of
  exposition, we only consider cases where a component has a
  single output interface (as do all of our example components).}
SL$_2$ and SL$_4$ are different labels for the same output interface.
If no inference rules apply, we show the preservation of input stream labels by applying
a default rule labeled ``(p).''

\begin{figure}
\begin{align*}
\text{protected}(\textbf{NDRead}_{gate}) \equiv \forall l \in \texttt{Labels  }  l = \textbf{NDRead}_{gate} \lor \\
\exists key \text{  } l = \textbf{Seal}_{key} \land \text{compatible}(gate, key)
\end{align*}
\begin{prooftree}
\begin{prooftree}
\AxiomC{\textbf{Taint} $\in$ \texttt{Labels}}
\UnaryInfC{\emph{Rep} ? \textbf{Diverge} : \textbf{Run}}
\end{prooftree}

\AxiomC{$\exists gate \exists \textbf{NDRead}_{gate} \in \texttt{Labels }$}
                          \AxiomC{$\lnot \text{protected}(\textbf{NDRead}_{gate})$}
\BinaryInfC{\emph{Rep} ? \textbf{Inst} : \textbf{Run}}
\end{prooftree}

\caption{The \emph{reconciliation} procedure applies the rules above to the set \texttt{Labels}, possibly adding additional labels.
``\emph{Rep} ? \textbf{A} : \textbf{B}'' means `if \emph{Rep}, add \textbf{A} to
\texttt{Labels}, otherwise add \textbf{B}.'
Finally,
\emph{reconciliation} returns the elements in \texttt{Labels} with highest severity.
}
\label{fig:reconciliation}
\end{figure}

\subsection{Coordination Selection}
\label{sec:coordination}

\blazes will automatically repair
dataflows that are not confluent or convergent by constraining how messages are delivered to
certain components.
When possible, \blazes will recognize the compatibility between sealed streams and component semantics,
synthesizing a seal-based strategy that avoids global coordination.  Otherwise, it will
enforce a total order on message delivery to those components.

\subsubsection{Sealing Strategies}

If the programmer has provided a seal annotation \textbf{Seal$_{key}$} that is compatible with the (non-confluent) component annotation,
we may use a synchronization strategy that avoids global coordination.
The intuition is that if the component never combines inputs from different (punctuated) partitions, then the order
in which it learns about the partitions, their contents and their corresponding seals has no effect on its outputs.
Consider a component representing a reporting server executing the query \emph{WINDOW} from Section~\ref{sec:example}.
Its label is $OR_{id, window}$.  
%Recall that a label of \textbf{Seal$_{window}$} on the click input stream represents 
%a contract that the stream
%partitions associated with different values of the windows attribute have the property that they ``stop changing.''
%We can see that the predicate of the first rule in Figure~\ref{fig:seal-rules} is satisfied.  
We know that \emph{WINDOW} will produce deterministic output contents if we delay its execution until we have accumulated
a complete, immutable partition to process (for a given value of the \emph{window} attribute). 
Thus a satisfactory protocol must allow stream producers to communicate
when a stream partition is sealed and what it contains, so that consumers can determine 
when the complete contents of a partition are known.

To determine that the complete partition contents are available, the consumer must a) participate in a protocol with 
each producer to ensure that the local per-producer partition is complete, and b) perform a unanimous voting protocol to 
ensure that it has received partition data from each producer.
Note that the voting protocol is a local form of one-way coordination, limited to the ``stakeholders'' contributing to
or consuming individual stream partitions.
When there is only one producer instance per partition, \blazes need not synthesize a voting protocol.  

Once the consumer has determined that the contents of a partition are 
immutable, it may process the partition without any further
synchronization.

\subsubsection{Ordering Strategies}
\label{sec:ordering_abstract}

If sealing strategies are not available,
\blazes achieves convergence for replicated, non-confluent components by using an
ordering service to ensure that all replicas process state-modifying events in
the same order.
%\blazes can utilize any totally ordered messaging service.
Our
current prototype uses a totally ordered messaging service based on Zookeeper for Bloom programs; for Storm, we use Storm's
built-in support for ``transactional'' topologies, which enforces a total order
over commits.

\section{Case studies}
\label{sec:case}

In this section, we apply \blazes to the examples introduced in
Section~\ref{sec:example}.  We describe how programmers can manually annotate
dataflow components.  We then discuss how \blazes identifies the coordination
requirements and, where relevant, the appropriate locations in these programs
for coordination placement.  
In Section~\ref{sec:eval} we will show concrete performance benefits of the \blazes
coordination choices as compared to a conservative use of a coordination service
such as Zookeeper.

We implemented the Storm wordcount dataflow, which consists of three ``bolts''
(components) and two distinct ``spouts'' (stream sources, which differ for the
coordinated and uncoordinated implementations) in roughly 400 lines of Java.  We
extracted the dataflow metadata from Storm into \blazes via a reusable adapter;
we describe below the output that \blazes produced and the annotations we added manually.  
We implemented the ad
reporting system entirely in Bloom, in roughly 125 LOC.  As discussed in the
previous section, \blazes automatically extracted all the relevant annotations.

For each dataflow, we present excerpts from the \blazes configuration file, 
containing the programmer-supplied annotations.  The interested reader
should refer to the technical report~\cite{blazes-tr} for details
on the derivation of each output stream label using the \blazes analyzer.

\subsection{Storm wordcount}

We first consider the Storm distributed wordcount query.  Given proper dataflow annotations, 
\blazes indicates that global ordering of computation on different components
is unnecessary to ensure deterministic replay, and hence consistent outcomes.

\subsubsection{Component annotations}
\label{sec:case-storm}

To annotate the three components of the Storm word count query, we provide the following file
to \blazes:

\begin{alltt}
\begin{scriptsize}
  Splitter:
    annotation:
      - \{ from: tweets, to: words, label: CR \}
  Count:
    annotation:
      - \{ from: words, to: counts, label: OW,
        subscript: [word, batch] \}
  Commit:
    annotation: \{ from: counts, to: db, label: CW \}
\end{scriptsize}
\end{alltt}

\texttt{Splitter}
is a stateless, confluent component: we
give it the annotation $CR$. 
We annotate \texttt{Count} as $OW_{word, batch}$---it is stateful
(accumulating counts over time) and order-sensitive, but potentially sealable on
word or batch (or both).
Lastly, \texttt{Commit} is also stateful (the backing store to which it stores the 
final counts is persistent), but since it is append-only and does not record the order 
of appends, we annotate it $CW$.

\subsubsection{Analysis}

In the absence of any seal annotations, \blazes derives an output label
of \textbf{Run} for the wordcount dataflow:

\begin{scriptsize}
\begin{prooftree}
\def\defaultHypSeparation{\hskip .03in}
\AxiomC{\textbf{Async}}
\AxiomC{CR}
\LeftLabel{(p)}
\BinaryInfC{\textbf{Async}}
\LeftLabel{\small \texttt{Splitter}}
\UnaryInfC{\textbf{Async}}
                             \AxiomC{OW$_{word, batch}$}
                \LeftLabel{\small (2)}
                \BinaryInfC{Taint}
                \LeftLabel{\small \texttt{Count}}
                \UnaryInfC{\textbf{Run}}
                                      \AxiomC{CW}
                          \LeftLabel{(p)}
                          \BinaryInfC{\textbf{Run}}
                          \LeftLabel{\small \texttt{Committer}}
                          \UnaryInfC{\textbf{Run}}
\end{prooftree}

\end{scriptsize}

Without coordination, nondeterministic input orders may produce nondeterministic output contents due to the order-sensitive nature of the \texttt{Count}
component.  To ensure that replay (Storm's
internal fault-tolerance strategy) is deterministic,
\blazes will recommend that the topology be coordinated---the programmer can achieve this by making
the topology ``transactional'' (in Storm terminology), totally ordering the batch commits.

If, on the other hand, the input stream is sealed on \emph{batch}, \blazes 
recognizes the compatibility between the stream punctuations and the
\texttt{Count} component, which operates over grouping sets of
\emph{word, batch}:

\begin{scriptsize}
\begin{prooftree}
\def\defaultHypSeparation{\hskip .03in}
\AxiomC{\textbf{Seal$_{batch}$}}
\AxiomC{CR}
\LeftLabel{(p)}
\BinaryInfC{\textbf{Seal$_{batch}$}}
\LeftLabel{\small \texttt{Splitter}}
\UnaryInfC{\textbf{Seal$_{batch}$}}
                             \AxiomC{OW$_{word, batch}$}
                \LeftLabel{\small (p)}
                \BinaryInfC{\textbf{Async}}
                \LeftLabel{\small \texttt{Count}}
                \UnaryInfC{\textbf{Async}}
                                      \AxiomC{CW}
                          \LeftLabel{(p)}
                          \BinaryInfC{\textbf{Async}}
                          \LeftLabel{\small \texttt{Committer}}
                          \UnaryInfC{\textbf{Async}}

\end{prooftree}
\end{scriptsize}

Because a batch is atomic
(its contents may be completely determined once a seal record arrives) and independent (emitting
a processed batch never affects any other batches), the topology will produce deterministic 
outputs
%%---a requirement for Storm's replay-based fault-tolerance---
under all interleavings.

\subsection{Ad-reporting system}
Next we describe how we might annotate the various components of the ad-reporting system.  As we discuss in Section~\ref{sec:white},
these annotations can be automatically extracted from the Bloom syntax; for exposition, in this section we discuss how a programmer 
might manually annotate an analogous dataflow written in a language without Bloom's static-analysis capabilities.
As we will see, ensuring deterministic outputs will require different mechanisms for the 
different queries listed in Figure~\ref{fig:sql}.

\subsubsection{Component annotations}
\label{sec:case-ads}

Below is the \blazes annotation file for the ad serving network:

\begin{alltt}
\begin{scriptsize}
  Cache:
    annotation:
      - \{ from: request, to: response, label: CR \}
      - \{ from: response, to: response, label: CW \}
      - \{ from: request, to: request, label: CR \}
  Report:
    Rep: true
    annotation:
      - \{ from: click, to: response, label: CW \}
  POOR: \{ from: request, to: response, label: OR,
        subscript: [id] \}
  THRESH: \{ from: request, to: response, label: CR \}
  WINDOW: \{ from: request, to: response, label: OR,
        subscript: [id, window] \}
  CAMPAIGN: \{ from: request, to: response, label: OR,
        subscript: [id, campaign] \}
\end{scriptsize}
\end{alltt}

The cache is clearly a stateful component, but since its state is
append-only and order-independent
%~\footnote{This extremely simple cache provides no facility for
%deletion or modification of cache entries.  In Section~\ref{sec:conclusion}, 
%we will discuss the problem of state management
%for append-only components.}
we may annotate it $CW$.
Because the data-collection path through the reporting server simply appends clicks and impressions to a log,
we annotate this path $CW$ also.  

All that remains is to annotate the read-only path through the reporting component
corresponding to the various continuous queries listed in
%Section~\ref{sec:queries}.  
Figure~\ref{fig:sql}.
\texttt{Report} is a replicated component, so we supply the \textbf{Rep}
annotation for all instances.
We annotate the query path corresponding to
\emph{THRESH}---which 
is confluent because it never emits a record until the ad impressions
reach the given threshold---$CR$.
We annotate queries \emph{POOR} and \emph{CAMPAIGN} \textbf{$OR_{id}$}
and \textbf{$OR_{id, campaign}$}, respectively.
These queries can return different contents in different executions, recording the effect of message races between
click and request messages.
We give query \emph{WINDOW} the annotation \textbf{$OR_{id, window}$}.
Unlike \emph{POOR} and \emph{CAMPAIGN}, \emph{WINDOW} includes the input stream attribute \emph{window}
in its grouping clause.
Its outputs are therefore partitioned by values of \emph{window}, 
making it compatible with an input stream sealed on \emph{window}.

%so \blazes will be able to employ 
%a coordination-free
%sealing strategy to force the component to output deterministic results 
%if it can determine that the input stream is sealed
%on \emph{window}.
%As we argued, \emph{campaign} presents an alternative seal key that is logical rather than explicitly temporal in nature.
%Because campaigns are long-lived, waiting until a campaign is over before looking at how advertisements within a campaign
%perform may not be useful. \jmh{This seems like a point for a Discussion toward the end -- partition selection seems like a new, open design problem.}
%%Finally, query \emph{ALLWINDOW} is the same as queries \emph{POOR} and \emph{CAMPAIGN}; it arises when an input stream
%%sealed on \emph{window} flows into the \textbf{$R_{campaign}$} annotation.

\subsubsection{Analysis}

Having annotated all of the instances of the reporting server component for different queries, 
we may now consider 
how \blazes automatically derives output stream labels for the global dataflow.
If we supply \emph{THRESH}, \blazes 
derives a final label of \textbf{Async} for the 
output path from cache to sink:

\begin{scriptsize}
\begin{prooftree}
\def\defaultHypSeparation{\hskip .3in}
\AxiomC{\textbf{Async}}
\AxiomC{$CW$}
\LeftLabel{(p)}
\BinaryInfC{\textbf{Async}}
                    \AxiomC{\textbf{Async}}
                    \AxiomC{$CW$}
                    \LeftLabel{(p)}
                    \BinaryInfC{\textbf{Async}}
                                                \AxiomC{$Rep$}
        \LeftLabel{\small \texttt{Report}}

\def\defaultHypSeparation{\hskip .03in}
        \TrinaryInfC{\textbf{Async}}
                              \AxiomC{$CW$}
                    \LeftLabel{(p)}
                    \BinaryInfC{\textbf{Async}}
%                                        \AxiomC{$Rep$}
        \LeftLabel{\small \texttt{Cache}}
        \UnaryInfC{\textbf{Async}}

\end{prooftree}
\end{scriptsize}

All components are confluent, so
the complete dataflow produces deterministic outputs without coordination.  If we chose, we could encapsulate the service 
as a single component with annotation $CW$.

Given query \emph{POOR} with no input stream annotations, \blazes
derives a label of \textbf{Diverge}:

\begin{scriptsize}
\begin{prooftree}
\def\defaultHypSeparation{\hskip .3in}
\AxiomC{\textbf{Async}}
\AxiomC{$CW$}
\LeftLabel{(p)}
\BinaryInfC{\textbf{Async}}

                    \AxiomC{\textbf{Async}}
                    \AxiomC{OR$_{campaign}$}
                    \LeftLabel{(2)}
                    \BinaryInfC{\textbf{NDRead$_{campaign}$}}
                                                \AxiomC{$Rep$}
        \LeftLabel{\small \texttt{Report}}

\def\defaultHypSeparation{\hskip .03in}
        \TrinaryInfC{\textbf{Inst}}
                              \AxiomC{$CW$}
                    \LeftLabel{(3)}
                    \BinaryInfC{\textbf{Taint}}
%                                        \AxiomC{$Rep$}
        \LeftLabel{\small \texttt{Cache}}
        \UnaryInfC{\textbf{Split}}

\end{prooftree}
\end{scriptsize}

The poor performers query is not confluent: it produces
nondeterministic output contents.  Because these outputs mutate a stateful, replicated component 
(i.e., the cache) that affects system outputs,
the output stream is tainted by divergent replica state.
Preventing replica divergence will require a coordination strategy that
controls message delivery order to the reporting server.

If, however, the input stream is sealed on
\emph{campaign}, \blazes recognizes the compatibility between
the stream partitioning and the component path annotation $OR_{id,campaign}$,
synthesizes a protocol that allows the partition to be processed when it has
stopped changing, and gives the dataflow the label \textbf{Async}:

\begin{scriptsize}
\begin{prooftree}
\def\defaultHypSeparation{\hskip .03in}
\AxiomC{\textbf{Seal$_{campaign}$}}
\AxiomC{$CW$}
\LeftLabel{(p)}
\BinaryInfC{\textbf{Seal$_{campaign}$}}

                    \AxiomC{\textbf{Async}}
                    \AxiomC{OR$_{campaign}$}
                    \LeftLabel{(2)}
                    \BinaryInfC{\textbf{NDRead$_{campaign}$}}
                                                \AxiomC{$Rep$}
        \def\defaultHypSeparation{\hskip .03in} \LeftLabel{\small \texttt{Report}}
        \TrinaryInfC{\textbf{Async}}
                              \AxiomC{$CW$}
 \LeftLabel{(p)}
                    \BinaryInfC{\textbf{Async}}
%                                        \AxiomC{$Rep$}
        \LeftLabel{\small \texttt{Cache}}
        \UnaryInfC{\textbf{Async}}

\end{prooftree}
\end{scriptsize}

%Appropriately sealing inputs to
%non-confluent components can make them behave like confluent components.
Implementing this sealing strategy does not require global coordination,
but merely some synchronization between stream producers and consumers.

Similarly, \emph{WINDOW} (given an input stream sealed on \emph{window})
reduces to \textbf{Async}.

\section{Bloom Integration}
\label{sec:white}

To provide input for the ``grey box'' functionality of \blazes, programmers must convert
their intuitions about component behavior and execution topology into the annotations introduced
in Section~\ref{sec:grey}.  
As we saw in Section~\ref{sec:case-storm}, this process is often quite natural;
unfortunately, as we learned in Section~\ref{sec:case-ads}, it becomes increasingly burdensome as component
complexity increases.

Given an appropriately constrained language, the necessary annotations can be extracted automatically via static analysis.
In this section, we describe how we used the Bloom language to enable a transparent ``white box'' system, in which 
unadorned programs are submitted, analyzed and---if necessary to ensure consistent outcomes---automatically rewritten.
By applying techniques from database theory and logic programming, \blazes and Bloom allow programmers to shift their
focus from individual component behaviors to program outcomes---a significant step towards truly declarative programming for
distributed systems.

\subsection{Bloom components}

Bloom programs are bundles of declarative \emph{rules} describing the contents
of logical \emph{collections} and how they change over time.
To enable encapsulation and reuse, a Bloom program may
be expressed as a collection of \emph{modules} with
input and output interfaces
associated with relational schemas.
Hence modules map naturally to dataflow components.

Each module also defines an internal dataflow from input to output interfaces,
whose components are the individual rules.
\blazes analyzes this dataflow graph to automatically derive component annotations for Bloom modules.

\subsection{White box requirements}

To select appropriate component labels, \blazes needs to determine whether a component is confluent and whether
it has internal state that evolves over time.  To determine when sealing strategies are applicable, \blazes
needs a way to ``chase''~\cite{chase} the injective functional dependencies described in Section~\ref{sec:analysis} transitively
across compositions.

\subsubsection{Confluence and state}

As we described in Section~\ref{sec:calm}, the CALM theorem establishes that all \emph{monotonic} programs are
confluent.  The Bloom runtime includes analysis capabilities to identify---at the
granularity of program statements---nonmonotonic operations, which can be conservatively identified with a syntactic test.  
Any component free of such operations is provably order-insensitive.
Similarly, Bloom's type system distinguishes syntactically between transient event streams and stored tables.  A simple flow analysis automatically
determines if a component accumulates state over time.
Together, these analyses are sufficient to determine annotations (except for the subscripts, which we describe next)
for every Bloom statement in a given module.

\subsubsection{Support for sealing}

What remains is to determine the appropriate partition subscripts for non-confluent labels (the $gate$ in $OW_{gate}$ and $OR_{gate}$)
and to define an effectively computable procedure for detecting injective functional dependencies.

Recall that in Section~\ref{sec:comp-annotations} we chose a subscript for the SQL-like \texttt{WINDOW} query by considering its \emph{group by} clause;
by definition, grouping sets are independent of each other.
Similarly, the columns referenced in the \emph{where} clause of an antijoin identify sealable partitions.\footnote{\footnotesize To see this, note that we can deterministically
evaluate \texttt{select * from R where x not in (select x from S where y = `Yahoo!')} 
for any tuples of R once we have established that a.) there will be no more records in S with y = `Yahoo!', or b.) there will \emph{never} be a corresponding S.x.}
Applying this reasoning, \blazes selects subscripts in the following way:

\begin{enumerate}
  \item If the Bloom statement is an aggregation (\emph{group by}), the subscript is the set of grouping columns.
  \item If the statement is an antijoin (\emph{not in}), the subscript is the set of columns occurring in the theta clause.
\end{enumerate}

We can track the lineage of an individual attribute (processed by a nonmonotonic operator) by querying
Bloom's system catalog, which details how each rule application transforms (or
preserves) attribute values that appear in the module's input interfaces.
To detect injective functional dependencies (in a sound but incomplete way), 
we exploit the common special case
that the identity function is injective, as is any series of transitive
applications of the identity function.  For example, given $S \equiv
\pi_a\pi_{ab}\pi_{abc}R$, $S.a$ is injectively functionally determined by $R.a$.
%Second, programmers will often be able to indicate, in the form of annotations,
%that certain paths from component input attributes to output attributes
%represent injective functions.

%\footnote{It may also be the case that
%    programmers can indicate certain such paths as \emph{approximately}
%    injective.  For example, a programmer may ignore unlikely collisions and say
%    that the MD5 function is injective.}

\section{Evaluation}
\label{sec:eval}

In Section~\ref{sec:properties}, we considered the \emph{consequences of under-coordinating} distributed
dataflows.  
In this section, we measure the \emph{costs of over-coordination} by comparing the performance of two 
distinct dataflow systems, each under two coordination regimes: a generic order-based coordination strategy and
an application-specific sealing strategy.

We ran our experiments on Amazon EC2. In all cases, we average results over three runs; error bars are shown on the graphs.

\subsection{Storm wordcount}
%\nrc{Concerns: 1. I'd like to see end-to-end latency or throughput-at-saturation
%  as a function of cluster size; throughput over time is not very useful. 2. Did
%  you use the same settings for both variants? e.g., batch sizes. 3. What was the
%  input rate?}

To evaluate the potential savings of avoiding unnecessary synchronization in Storm, we
implemented two versions of the streaming wordcount query described in
Section~\ref{sec:example}.  Both process an identical stream of tweets and
produce the same outputs.  They differ in that the first implementation is a
``transactional topology,'' in which the \texttt{Commit} components 
coordinate to ensure that outputs are committed to the backing store in a
serial order.\footnote{Storm uses Zookeeper for coordination.}  The
second---which \blazes has ensured will produce deterministic outcomes without
any global coordination---is a ``nontransactional topology.'' We optimized the
batch size and cluster configurations of both implementations to maximize
throughput.

We used a single dedicated node (as the documentation recommends) 
for the Storm master and three
Zookeeper servers.  In each experiment, we allowed the topology to ``warm up''
and reach steady state by running it for 10 minutes.

\begin{figure}[t]
\centering
\includegraphics[width=0.8\linewidth]{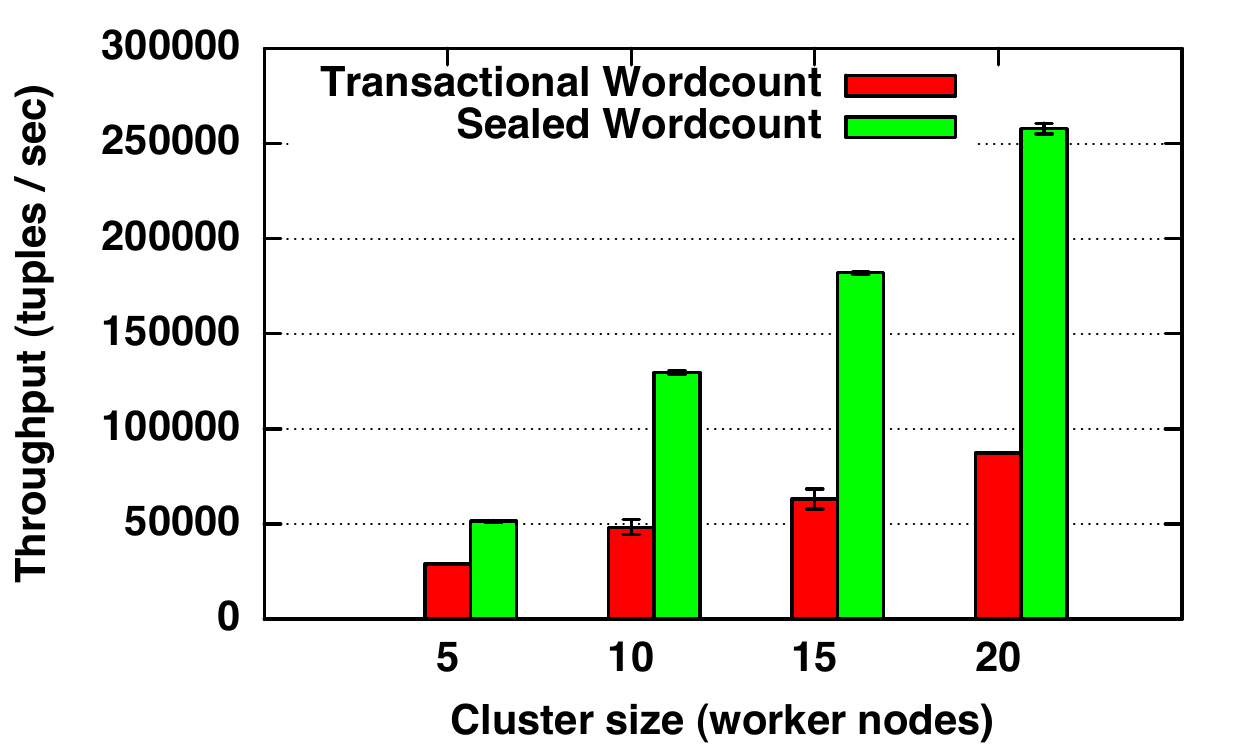}
\caption{\small The effect of coordination on throughput for a Storm topology computing
  a streaming wordcount.}
\label{fig:storm_eval}
\end{figure}

Figure~\ref{fig:storm_eval} plots the throughput of the coordinated and uncoordinated implementations of the wordcount dataflow
as a function of the cluster size.
The overhead of conservatively deploying a transactional topology is considerable.
The uncoordinated dataflow has a peak throughput roughly 1.8 times that of its coordinated counterpart in a 5-node deployment.
As we scale up the cluster to 20 nodes, the difference in throughput grows to 
$3 \times$.

\subsection{Ad reporting}

To compare the performance of the sealing and ordering coordination strategies, we conducted a series of
experiments using a \lang implementation of the ad tracking network introduced in Section~\ref{sec:example}.
For ad servers, which simply generate click logs and forward them to reporting
servers, we used 10 \texttt{micro} instances.  We created 3 reporting servers using \texttt{medium} instances.
Our Zookeeper cluster consisted of 3 \texttt{small} instances.  
%All instances were located in the same AWS availability zone. 
%\jmh{You had a single ZK instance in the storm case.  Even though you don't compare with this, it creates cognitive dissonance -- which is the ``right'' way?}

Ad servers generate a workload of 1000 log entries per server, dispatching 
50 click log messages in batch and sleeping periodically.  
During the workload, we pose a number of requests to the reporting servers,
all of which implement the continuous query \emph{CAMPAIGN}.

Although this system---implemented in the Bloom language prototype---does not illustrate the volume
we would expect in a high-performance implementation, we will see that it highlights some important \emph{relative}
patterns across different coordination strategies.

\subsubsection{Baseline: No Coordination}

For the first run, we do not enable the \blazes preprocessor.  Thus click logs and requests flow in an 
uncoordinated fashion to the reporting
servers.  The uncoordinated run provides a lower bound for performance of appropriately coordinated implementations.
However, it does not have the same semantics.  We confirmed by observation that 
%the reporting servers 
%do not return
%consistent results for the set of queries posed to all servers.
certain queries posed to multiple reporting server replicas returned inconsistent results.
The line labeled ``Uncoordinated'' in Figures~\ref{fig:5srv} and~\ref{fig:10srv} shows the log records processed
over time for the uncoordinated run, for systems with 5 and 10 ad servers, respectively.

\subsubsection{Ordering Strategy}

In the next run we enabled the \blazes preprocessor but did not supply any input
stream annotations.  \blazes recognized the potential for inconsistent answers
across replicas and synthesized a coordination strategy based on ordering.  By
inserting calls to Zookeeper, all click log entries and requests were delivered
in the same order to all replicas.  The line labeled ``Ordered'' in
Figures~\ref{fig:5srv} and~\ref{fig:10srv} plots the records processed over time for this
strategy.

The ordering strategy ruled out inconsistent answers from replicas but incurred a significant performance penalty.
Scaling up the number of ad servers by a factor of two had little effect on the performance
of the uncoordinated implementation, but increased the processing time in the coordinated run by
a factor of three.

\subsubsection{Sealing Strategies}
\label{sec:seal-strat}

For the last experiments we provided the input annotation
$\textbf{Seal}_{campaign}$ and embedded punctuations in the ad click stream
indicating when there would be no further log records for a particular campaign.
Recognizing the compatibility between the sealed stream and the
aggregate query in \emph{CAMPAIGN} (a ``group-by'' on \emph{id, campaign}),
\blazes synthesized a seal-based coordination strategy.

Using the seal-based strategy, reporting servers do not
need to wait until events are globally ordered;
instead, they are processed as soon as a reporting server can determine that
they belong to a sealed partition.  
%After each ad server forwards
%its final click record for a given campaign to the replicated reporting servers,
%it sends a seal message for that campaign, which contains a digest of the set of
%click messages it generated.  
The reporting servers use Zookeeper
only to determine the set of ad servers responsible for each campaign---that is, one call to Zookeeper
per campaign.
%per partition rather than per click message.  
When a 
reporting server has
received seal messages from all producers for a given campaign, 
%it compares the
%buffered click records to the seal digest(s); if they match, 
it emits the partition for processing.

\begin{figure}[t]
\centering
\includegraphics[width=0.8\linewidth]{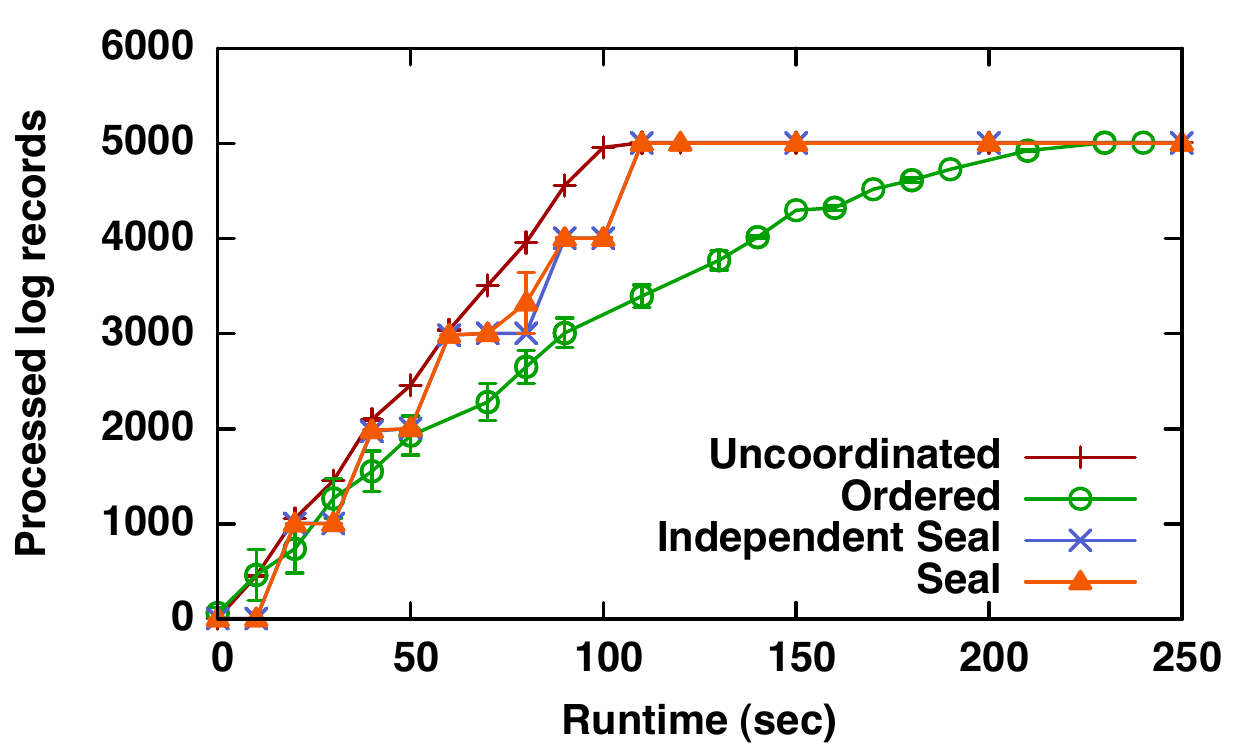}
\caption{\small Log records processed over time, 5 ad servers.}
\label{fig:5srv}
\end{figure}

\begin{figure}[t]
\centering
\includegraphics[width=0.8\linewidth]{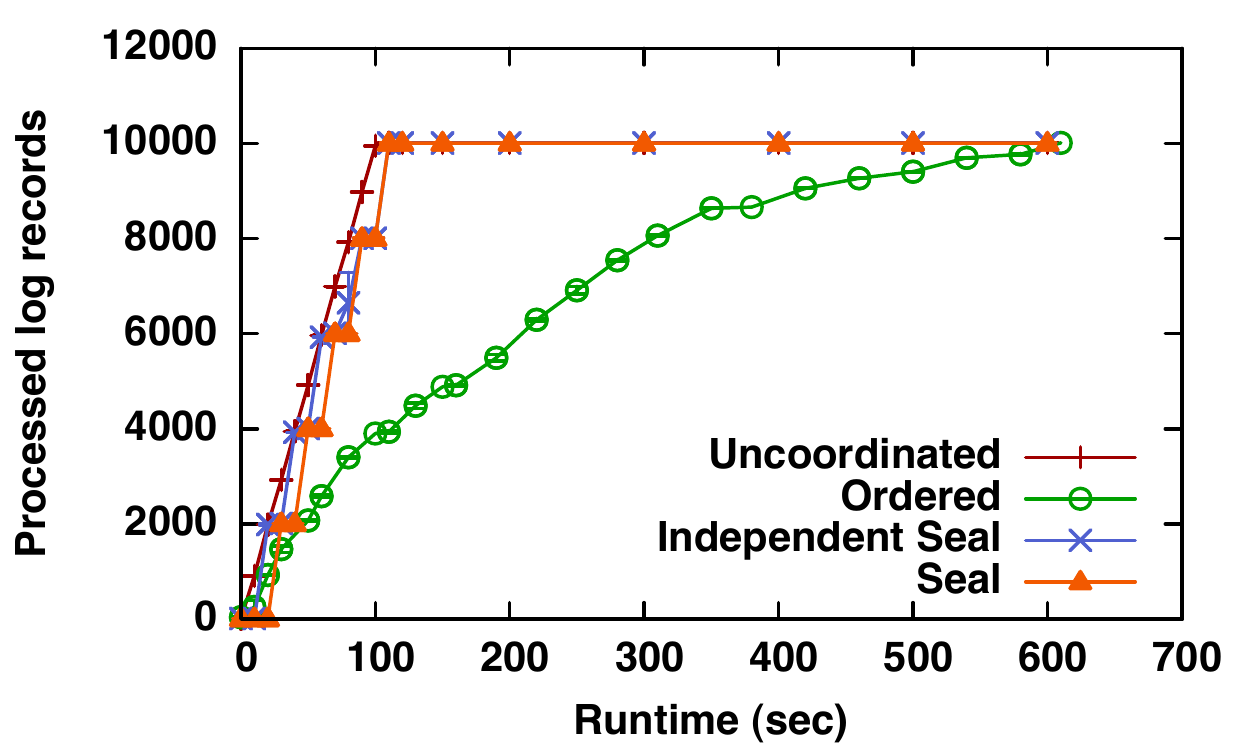}
\caption{\small Log records processed over time, 10 ad servers.}
\label{fig:10srv}
\end{figure}

%Figures~\ref{fig:5srv} and \ref{fig:10srv} compare the performance of seal-based
%strategies to ordered and uncoordinated runs.  
In Figures~\ref{fig:5srv} and \ref{fig:10srv} we evaluate the sealing
strategy for two alternative partitionings of click records:
in ``Independent seal'' each campaign is
mastered at exactly one adserver, while in ``Seal,'' all ad servers produce
click records for all campaigns.  Note that both seal-based runs closely
track the performance of the uncoordinated run; doubling the number of ad 
servers 
%has little effect on the system throughput.
effectively doubles system throughput.

\begin{figure}[t]
\centering
\includegraphics[width=0.8\linewidth]{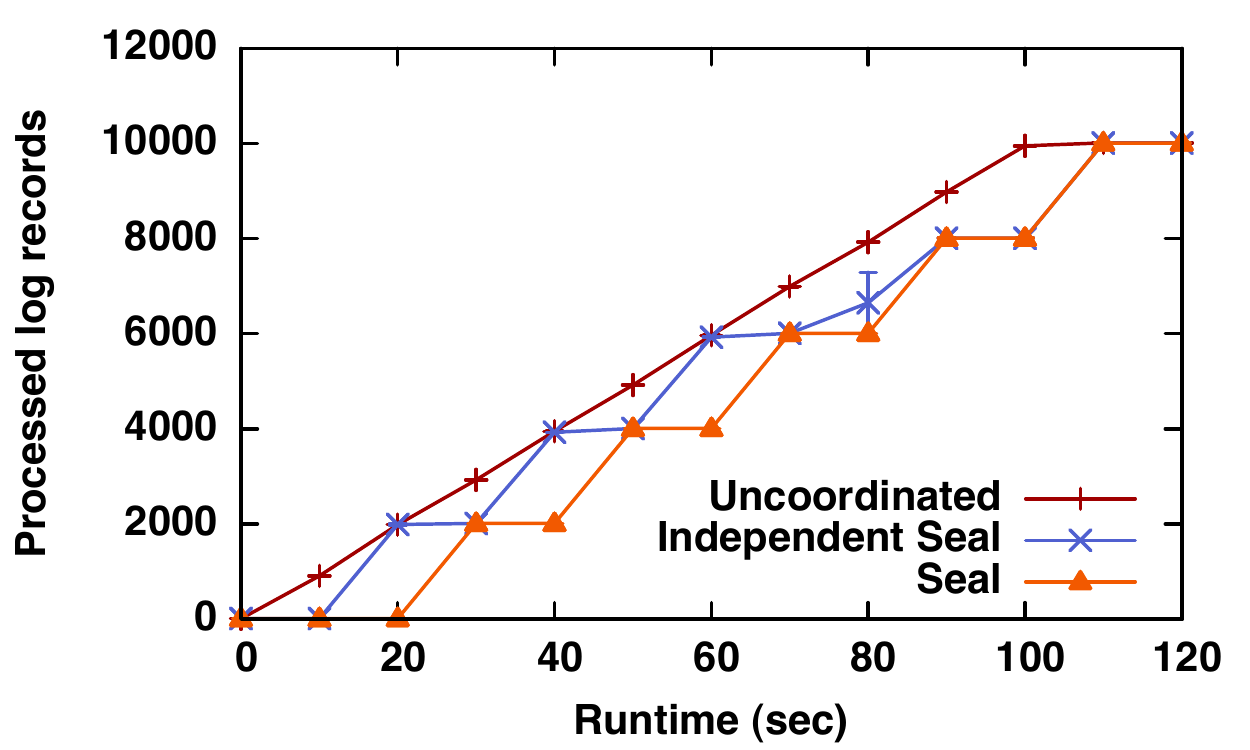}
\caption{\small Seal-based strategies, 10 ad servers.}
\label{fig:10seals}
\end{figure}

To highlight the differences between the two seal-based runs,
Figure~\ref{fig:10seals} plots the 10-server run but omits the ordering strategy.
As we would expect, ``independent seals'' result in
lower latencies because reporting servers may process partitions as soon as
a single seal message appears (since each partition has a single producer).
By contrast, the step-like shape of the non-independent seal strategy 
reflects the fact that reporting servers delay processing input partitions
until they have received a seal record from every producer.  
Partitioning the data across ad servers so as to place advertisement
content close to consumers (i.e., partitioning by ad id) caused campaigns to be spread across
ad servers, conflicting with the coordination strategy.
We revisit the notion of ``coordination locality'' in Section~\ref{sec:conclusion}.

\section{Related Work}

Our approach to automatically coordinating distributed services draws
inspiration from the literature on both distributed systems and databases.
Ensuring consistent replica state by establishing a total order of message
delivery is the technique adopted by state machine replication~\cite{statemachine};
each component implements a deterministic state machine, and a global coordination
service such as atomic broadcast or Multipaxos decides the message order.

%In the context of Dedalus,
%---a logic language for programming distributed systems
%that is closely related to Bloom---
Marczak et al.\ draw a connection between
stratified evaluation of conventional logic programming languages and
distributed protocols to ensure consistency~\cite{dedalus-model}.  They describe
a program rewrite that ensures deterministic executions by preventing any node
from performing a nonmonotonic operation until that operation's inputs 
have stopped changing.
%%---that is, henceforth unchanging.  
%Agents processing or contributing to a distributed relation carry out
%a voting-based protocol to agree when the contents of the relation are
%completely determined.  
This rewrite---essentially a restricted version of the
sealing construct defined in this paper---treats entire input collections as
sealable partitions,
%It ensures both deterministic replica state within an
%execution and deterministic results across executions, but 
and hence
is not defined for
unbounded input relations.

Commutativity of concurrent operations is a subject of interest for
parallel as well as distributed programming languages.  Commutativity analysis~\cite{commutativityanalysis} 
uses symbolic analysis to test whether different method-invocation orders always lead to the same result;
when they do, lock-free parallel executions are possible.  
%In the parallel functional language $\lambda_{\textrm{LVar}}$~\cite{Kuper2013},
%program state is constrained to grow according to a partial order and queries are restricted,
%enabling the creation of programs that are ``deterministic by construction.''
CRDTs~\cite{crdts} are convergent replicated data structures; CRDTs can be modeled in
\blazes as components whose update API calls are labeled $CW$.
LVar data structures~\cite{Kuper2013} ensure determinism for shared-memory
parallel programs by restricting modifications and observations of
shared state according to a user-specified lattice.
Like confluent components, CRDTs and LVars ensure monotone growth of state.

Like reactive distributed systems, streaming databases~\cite{aurora, stream, telegraphcq} 
must operate over unbounded inputs---we have borrowed much of our stream formalism from
this tradition.  The CQL language
distinguishes between monotonic and nonmonotonic operations; the former support efficient strategies for 
converting between streams and relations due to their pipelineability.  The Aurora system also distinguishes between
``order-agnostic'' and ``order-sensitive'' relational operators.  

Similarly to our work, the Gemini system~\cite{redblue} attempts to
efficiently and correctly evaluate a workload
with heterogeneous consistency requirements, ensuring replica convergence while
taking advantage of cheaper strategies for operations that 
require only weak orderings.  
By contrast, \blazes makes guarantees about composed services, which requires reasoning
about the properties of streams as well as component state.

\section{Conclusions}
\label{sec:conclusion}

\blazes relieves programmers of the burden of deciding \emph{when} and
\emph{how} to use the (precious) resource of distributed coordination.  With
this difficulty out of the way, the programmer may focus their insight on
other difficult problems, such as \emph{placement}---both the physical placement
of data and the logical placement of components.

Rules of thumb regarding data placement strategies typically involve predicting patterns of access that exhibit
spatial and temporal locality; data items that are accessed together should
be near one another, and data items accessed frequently should be cached.  Our discussion
of \blazes, particularly the evaluation of different seal-based strategies in Section~\ref{sec:seal-strat},
hints that access patterns are only part of the picture: because the dominant cost in large-scale systems
is distributed coordination, 
we must also consider \emph{coordination locality}---a rough measure being the number
of nodes that must communicate to deterministically process a segment of data. 
%If coordination locality is in conflict with spatial locality 
%(e.g., the non-independent partitioning strategy that clusters ads likely to be served together at the cost of distributing
%%campaigns across multiple nodes),
%problems emerge.
Problems emerge if coordination locality is in conflict with spatial locality.  For example, clustering ads
likely to be served together (the non-independent seal topology in Figure~\ref{fig:10seals}) caused campaigns (the 
seal key) to be distributed across multiple nodes, increasing coordination latency.

Given a dataflow, \blazes determines the need for (and appropriately applies)
coordination.  But was it the right dataflow?  We might wish to ask whether a
different logical dataflow that produces the same output supports cheaper
coordination strategies.  Some design patterns emerge from our discussion.  The
first is that, when possible, replication should be placed \emph{upstream} of
confluent components.  Since they are tolerant of all input orders, inexpensive
replication strategies (like gossip) are sufficient to ensure confluent outputs.
Similarly, caches should be placed \emph{downstream} of confluent components.
Since such components never retract outputs, simple, append-only caching logic
may be used.  More challenging and compelling is the possibility of capturing
these design principles into a compiler and automatically rewriting dataflows.

\section*{Acknowledgments}
We would like to thank
Michael Armbrust, Peter Bailis, Alan Fekete, Andy Gross, Coda Hale, Tim Kraska, Josh Rosen, Dmitriy Ryaboy
and the anonymous reviewers for their helpful feedback on this paper.
This work was supported by the Air Force Office of Scientific Research (grant
FA95500810352), the Natural Sciences and Engineering Research Council of Canada,
the National Science Foundation (grants CNS-0722077, IIS-0713661, IIS-0803690,
and IIS-0917349), and gifts from EMC, Microsoft Research, and NTT Multimedia
Communications Laboratories.

\bibliographystyle{IEEEtran}
\bibliography{icde14}

\end{document}